\documentclass[fleqn,usenatbib]{mnras}
\pdfoutput=1
\usepackage{newtxtext,newtxmath}
\usepackage[T1]{fontenc}
\usepackage[utf8]{inputenc}
\usepackage{ae,aecompl}

\usepackage[encapsulated]{CJK}	
\usepackage{graphicx}	
\usepackage{bm}
\usepackage{subfigure}
\usepackage[normalem]{ulem}

\usepackage{etoolbox}
\makeatletter
\patchcmd\@combinedblfloats{\box\@outputbox}{\unvbox\@outputbox}{}{\errmessage{\noexpand patch failed}}
\makeatother

\newcommand{\tder}[2]{\frac{\mathrm{d}#1}{\mathrm{d}#2}}  
\renewcommand{\vec}[1]{\bm{#1}}
\newcommand{\cntext}[1]{\begin{CJK}{UTF8}{bkai}#1\ignorespacesafterend\end{CJK}}  
\newcommand{\Hg}{H_\text{g}}  
\newcommand{\Mp}{M_\text{p}}  
\newcommand{\Mth}{M_\text{th}}  
\newcommand{\OmegaK}{\Omega_\text{K}}  
\newcommand{\RH}{R_\text{H}}  
\newcommand{\rs}{r_\text{s}}  
\newcommand{\cs}{c_\text{s}}  
\newcommand{\taus}{\tau_\text{s}}  
\newcommand{\pc}{\textsc{Pencil Code}}  
\newcommand{\ug}{\vec{u}_\text{g}}  
\newcommand{\vp}{\vec{v}_\text{p}}  
\newcommand{\Sigmag}{\Sigma_\text{g}}  
\newcommand{\Sigmap}{\Sigma_\text{p}}  

\title[Dusty discs with an embedded planet]
    {Morphological signatures induced by dust back reaction in discs with an embedded planet}

\author[C.-C.~Yang \& Z.~Zhu]{
Chao-Chin Yang (\cntext{楊朝欽})$^{1}$\thanks{E-mail: ccyang@unlv.edu}
and Zhaohuan Zhu (\cntext{朱照寰})$^{1}$
\\
$^{1}$Department of Physics and Astronomy,
    University of Nevada, Las Vegas,
    4505 S.~Maryland Parkway, Box~454002,\\
    Las Vegas, NV~89154-4002, USA
}

\date{Accepted 2019 November 13. Received 2019 November 8; in original form 2019 September 11}

\pubyear{2019}  

\begin{document}
\label{firstpage}
\pagerange{\pageref{firstpage}--\pageref{lastpage}}
\maketitle

\begin{abstract}
Recent observations have revealed a gallery of substructures in the dust component of nearby protoplanetary discs, including rings, gaps, spiral arms, and lopsided concentrations.
One interpretation of these substructures is the existence of embedded planets.
Not until recently, however, most of the modelling effort to interpret these observations ignored the dust back reaction to the gas.
In this work, we conduct local-shearing-sheet simulations for an isothermal, inviscid, non-self-gravitating, razor-thin dusty disc with a planet on a fixed circular orbit.
We systematically examine the parameter space spanned by planet mass ($0.1\Mth \le \Mp \le 1\Mth$, where $\Mth$ is the thermal mass), dimensionless stopping time ($10^{-3} \le \taus \le 1$), and solid abundance ($0 < Z \le 1$).
We find that when the dust particles are tightly coupled to the gas ($\taus < 0.1$), the spiral arms are less open and the gap driven by the planet becomes deeper with increasing $Z$, consistent with a reduced speed of sound in the approximation of a single dust-gas mixture.
By contrast, when the dust particles are marginally coupled ($0.1 \lesssim \taus \lesssim 1$), the spiral structure is insensitive to $Z$ and the gap structure in the gas can become significantly skewed and unidentifiable.
When the latter occurs, the pressure maximum radially outside of the planet is weakened or even extinguished, and hence dust filtration by a low-mass ($\Mp < \Mth$) planet could be reduced or eliminated.
Finally, we find that the gap edges where the dust particles are accumulated as well as the lopsided large-scale vortices driven by a massive planet, if any, are unstable, and they are broken into numerous small-scale dust-gas vortices.
\end{abstract}


\section{Introduction}

In the past decade, high-resolution observations of nearby protoplanetary discs have revealed a gallery of substructures in the distribution of dust particles.
Observing the scattered polarised light at the near infrared, the Strategic Explorations of Exoplanets and Disks with Subaru \citep[SEEDS;][]{mT16} survey collected an array of protoplanetary discs at a resolution of $\sim$10\,au, demonstrating morphologically diverse substructures like spiral arms, rings, and gaps.
Similar diversity in protoplanetary discs have also been found and categorised with the Very Large Telescope \citep[VLT;][]{GB17,GB18}.
Using the Atacama Large Millimeter/submillimeter Array (ALMA), \cite{MD13} detected a large-scale, lopsided concentration of dust particles in the disc around Oph~IRS~48, and \cite{AB15} showed that the dust distribution around HL~Tauri is almost perfectly axisymmetric, consisting of concentric rings and gaps.
Recently, the ALMA large program ``Disk Substructures at High Angular Resolution Project'' \citep[DSHARP;][]{AH18} published a homogeneous sample of 20 bright targets at a resolution of $\lesssim$5\,au and found that almost all discs show axisymmetric substructures in their dust distribution.
A similar result using ALMA was also found by \cite{LP18} in their sample of 12 resolved discs in the Taurus star-forming region.

A popular interpretation of the diverse substructures in the dust component of the observed protoplanetary discs is the existence of embedded young planets.
For example, the axisymmetric rings and gaps in the HL~Tauri disc can be explained with three planets of sub-Jovian masses (\citealt{DP15}; \citealt*{DZW15}).
Moreover, the five dust rings around AS~209 observed by DSHARP can even be explained by one single planet \citep{ZZ18}, a serendipitous case in which the relative locations and density contrasts match those of multiple dust rings that are simultaneously induced by one low-mass planet in a disc with low viscosity (\citealt{ZS14}; \citealt*{BZH17}; \citealt{DLC17,DLC18}).
The two-arm spirals as well as several other features observed in the discs of MWC~758 and SAO~206462 (HD~135344B) can be driven by a planetary companion (\citealt{DZ15}; \citealt*{BZH16}; \citealt*{DL18}).
Finally, large-scale vortices can be generated near the gap edges driven by a massive planet and entrain dust materials, leading to a lopsided structure around the star \citep[e.g.,][]{LJ09}, as in the observed disc of Oph~IRS~48 \citep[see also the case of HD~142527;][]{CP13}.

However, most of the modelling efforts to account for the observed substructures in nearby protoplanetary discs so far ignored the dust back reaction to the gas drag.
Meanwhile, it has become clear that this dust back reaction plays an active role in the dust-gas dynamics.
For instance, \cite{KM18} showed that a dust ring accumulated near the outer edge of a gap driven by a planet can be significantly broadened when both the dust back reaction and the dust diffusion are appreciable \citep*[see also][]{GLM17}.
By contrast, \cite{DL19} did not find such a broadening of dust ring with a full size distribution and evolution of dust materials.
Moreover, both \cite{CL19} and \cite*{PLR19} demonstrated that the dust back reaction can make a gap edge subject to the Rossby wave instability \citep{LL99}, leading to dust-gas vortices.
Therefore, further studies of the dust back reaction in the context of planet--disc interaction and its observational signatures seem warranted.

In this work, we focus our attention on the morphological signatures of the dust back reaction in both the gas and the dust components of a protoplanetary disc under the influence of an embedded planet on a fixed orbit, assuming the inviscid limit.
We conduct numerical simulations in 2D and systematically examine the parameter space spanned by the planet mass, the stopping time (or equivalently the size of the dust particles), and the abundance of the solids (or equivalently the strength of the back reaction), as described in Section~\ref{S:method}.
In Section~\ref{S:nbr}, we summarise the models in the limit of no dust back reaction as a baseline for comparison purposes.
We separate our discussion of the results with dust back reaction by models with tightly coupled dust particles in Section~\ref{SS:tcdp} and those with marginally coupled dust particles in Section~\ref{SS:mcdp}.
In Section~\ref{S:obs}, we briefly discuss the implications of our models for observations.
Finally, we conclude in Section~\ref{S:summary} with a summary of the effects we have observed in our models and some caveats.

\section{Methodology} \label{S:method}

We consider an isothermal, inviscid, razor-thin, non-self-gravitating circumstellar disc and adopt the classical local-shearing-sheet approximation \citep{GL65}.
The disc is mapped by a non-inertial polar coordinate system $(r, \theta)$ with its origin at the central star of mass $M_\star$.
The system rotates around the star with the Keplerian angular frequency $\OmegaK$ at a certain radial distance $r = r_0$ to the star.
Focusing on only a patch of the coordinate system centred at $(r_0, \theta_0)$, where $\theta_0$ is an arbitrary reference polar angle, a local shearing sheet defines a pair of new coordinates $x \equiv r - r_0$ and $y \equiv r(\theta - \theta_0)$ and linearises the equations of motion for the gas and the dust particles to first order in $x / r_0$ and $y / r_0$.
In this frame of reference, all velocities are measured with respect to the background Keplerian shear velocity $-3\OmegaK x \hat{\bm{e}}_y / 2$.

The gaseous and the dust components of the disc interacts via their mutual drag force.
For simplicity, we assume that in each model, the dust particles have the same dimensionless stopping time $\taus \equiv \OmegaK t_\text{s}$, where $t_\text{s}$ is the characteristic timescale for reducing the relative velocity between a dust particle and its surrounding gas by the drag force.
The value of $\taus$ depends on the properties of the particle and the gas, and usually the smaller $\taus$, the smaller the particle and/or the denser the gas \citep[see, e.g.,][]{JB14}.

To evolve the system of gas and dust, we use the \pc{} \citep{BD02}.
The gaseous component of the disc is modelled on a fixed regular grid, and is integrated with sixth-order spatial derivatives and a third-order Runge--Kutta method.
We stabilise the scheme by sixth-order hyper-diffusion with a fixed Reynolds number, which damps numerical noise near the Nyquist frequency while obtaining high fidelity over a large dynamical range \citep{YK12}.
Shocks are captured via artificial diffusion in every dynamical variable, and a correction term to the momentum equation is applied so that the total momentum is better conserved \citep{GM18}.
To relieve the Courant condition imposed by the background Keplerian shear velocity at large radial distances from the origin of the local shearing sheet, we use a sixth-order B-spline interpolation to integrate the shear advection terms \citep*{YJC17,YMJ18}.

The dust component of the disc is modelled as Lagrangian super-particles, each of which represents an ensemble of numerous identical dust particles.
The position and velocity of each super-particle is integrated simultaneously with the gas using the same Runge--Kutta steps.
The gas grid and the dust particles are coupled via their mutual drag force, and the computation of this coupling is achieved by the standard particle-mesh method with the Triangular-Shape-Cloud weighting scheme \citep{HE88}.
Because we consider a wide range of dimensionless stopping time $\taus$ with potentially strong dust concentration, the mutual drag force can be stiff and hence we use the numerical algorithm devised by \cite{YJ16} to integrate the force efficiently.

The gaseous disc usually has a radial pressure gradient on large scale.
Instead of initialising the gas with such a gradient, we impose it as a constant external source term.
The gradient is such that the azimuthal velocity of the gas is reduced by $\Delta u$ from the Keplerian velocity when the disc is free of dust.
In this work, we consider only the case that $\Delta u / \cs = 0.05$, where $\cs$ is the speed of sound.
This value is similar to that in the inner region ($\lesssim$10\,au) of typical protoplanetary disc models \cite[e.g.,][]{BJ15}.

\begin{figure*}
	\includegraphics[width=\textwidth]{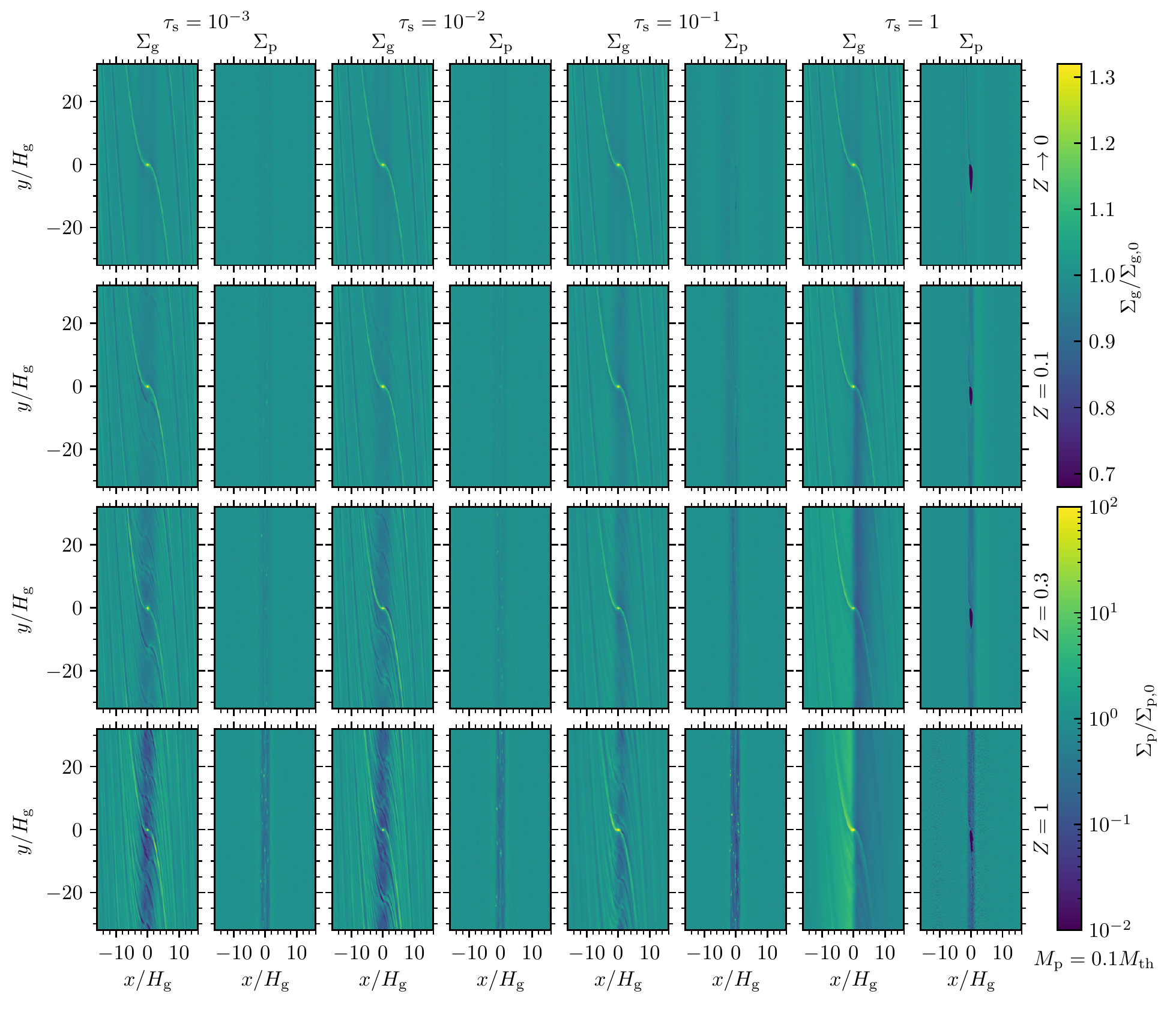}
    \caption{Column densities of the gas $\Sigmag$ and the particles $\Sigmap$ at the end of the simulations ($t = 100P$) from models with a planet of mass $\Mp = 0.1\Mth$, where $\Mth$ is the thermal mass (equation~\eqref{E:mth}).
        For each model, the gas and the dust are shown respectively on the left and on the right in a double panel.
        The panels are organised such that the dimensionless stopping time $\taus$ increases from left to right, while the solid abundance $Z$ increases from top to bottom.
        The densities are normalised by their initial values $\Sigma_{\text{g},0}$ and $\Sigma_{\text{p},0}$, respectively.}
    \label{F:sigma1}
\end{figure*}

\begin{figure*}
	\includegraphics[width=\textwidth]{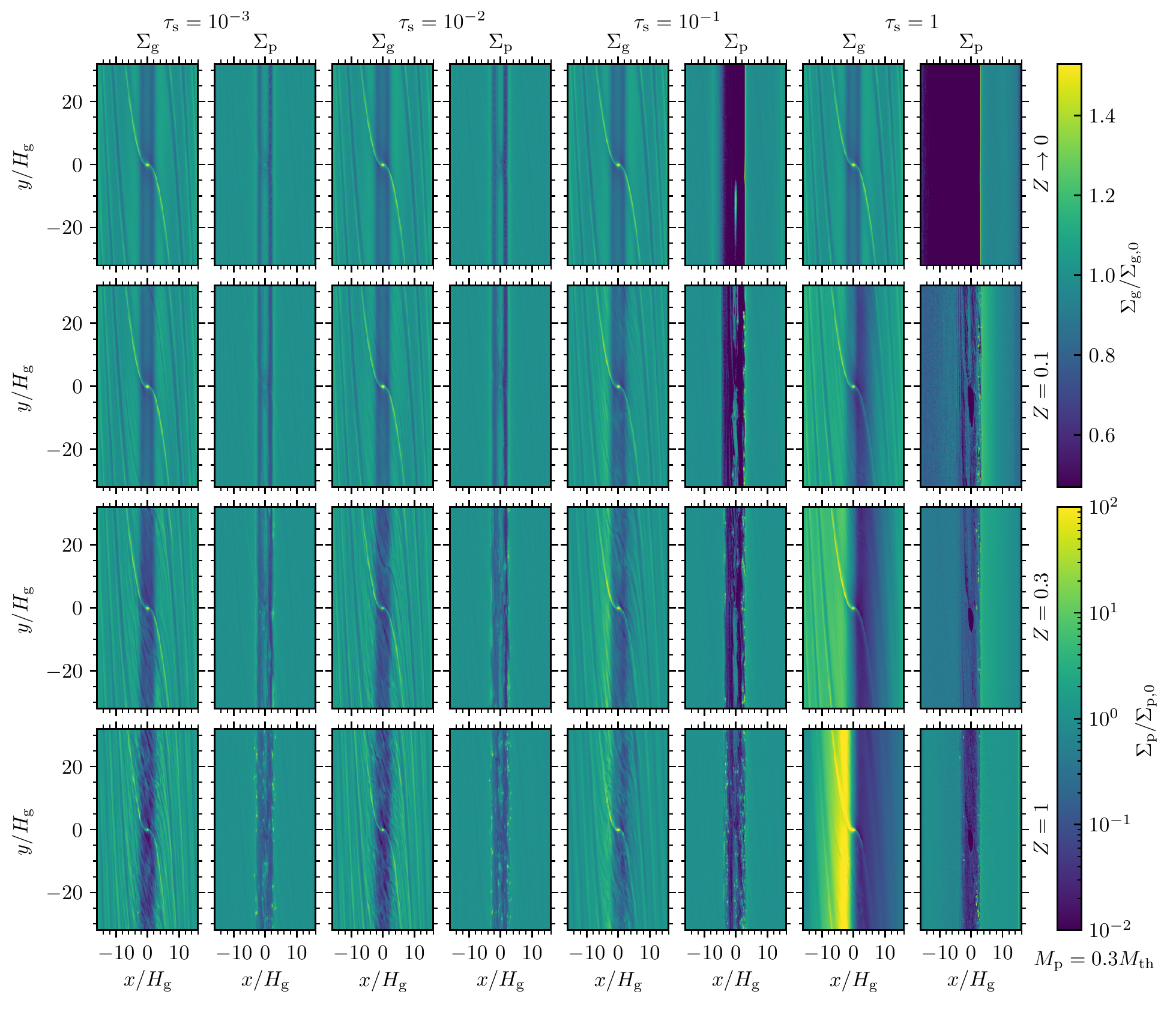}
    \caption{The same as Fig.~\ref{F:sigma1} except for models with $\Mp = 0.3\Mth$.}
    \label{F:sigma2}
\end{figure*}

\begin{figure*}
	\includegraphics[width=\textwidth]{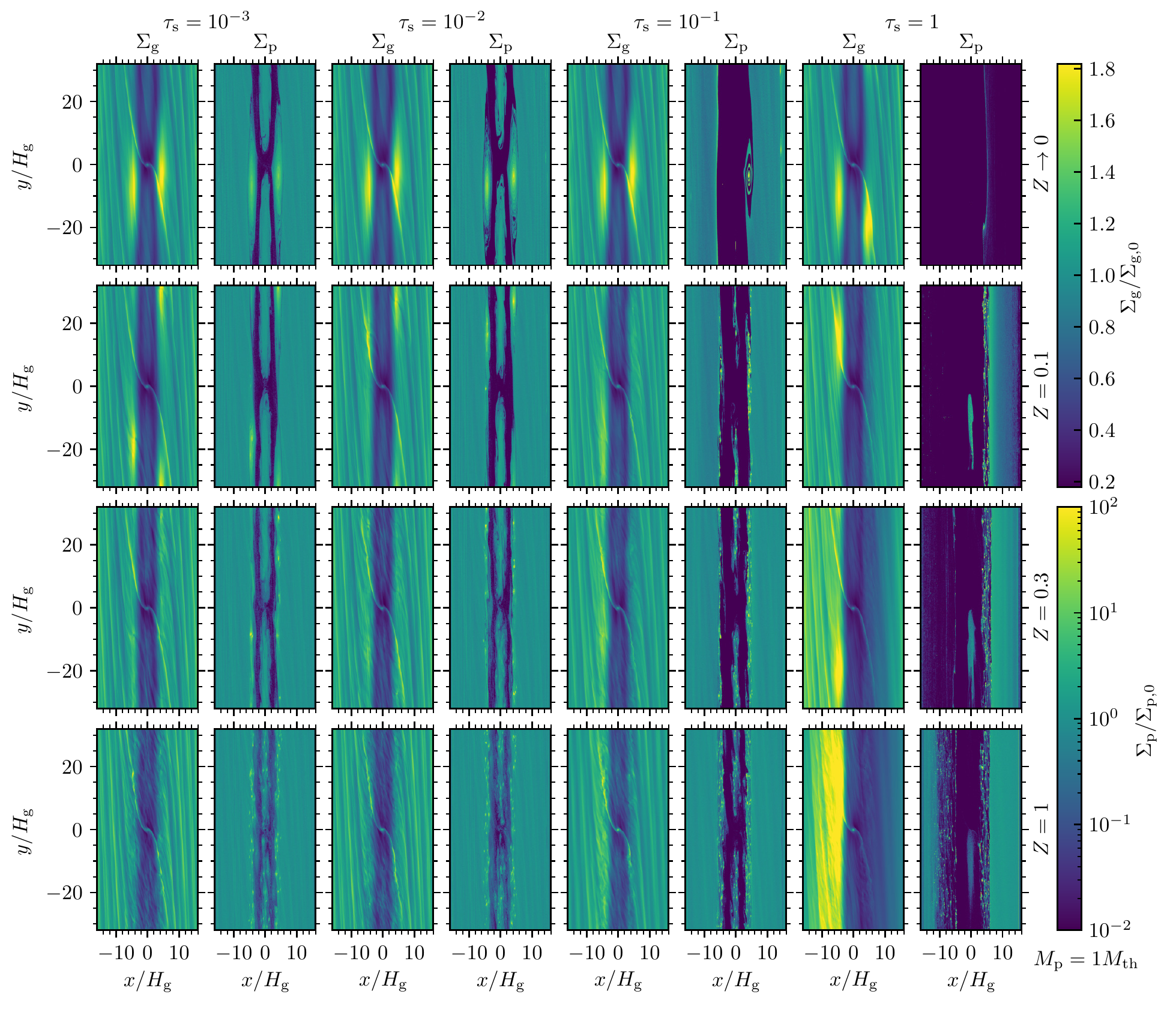}
    \caption{The same as Fig.~\ref{F:sigma1} except for models with $\Mp = 1\Mth$.}
    \label{F:sigma3}
\end{figure*}

We place a point-like planet of mass $\Mp$ fixed at the centre of the local shearing sheet.
To avoid the singularity of the gravitational force exerted by the planet, we adopt a smoothed gravitational potential of the form
\begin{equation} \label{E:gravpot}
\Phi(d) = -G\Mp\frac{d^2 + 3\rs^2 / 2}{(d^2 + \rs^2)^{3/2}},
\end{equation}
where $G$ is the gravitational constant, $d \equiv \sqrt{x^2 + y^2}$ is the distance to the planet, and $\rs$ is the smoothing length (\citealt{DR11}; \citealt*{ZSR12}).
For $d \gg r_s$, $\Phi(d) \simeq -(G\Mp / d)(1 - 9\rs^4 / 4d^4)$, while for $d \ll r_s$, $\Phi(d) \simeq \Phi_0 + 5G\Mp d^2 / 4\rs^3$, where $\Phi_0$ is a constant.
The planet mass $\Mp$ can be expressed in terms of the thermal mass \citep{GR01,DR11}
\begin{equation} \label{E:mth}
\Mth \equiv
\frac{\cs^3}{G\OmegaK} =
\left(0.096\,M_\text{J}\right)
\left(\frac{\cs}{0.6\,\text{km s}^{-1}}\right)^3
\left(\frac{M_\star}{M_\odot}\right)^{-1/2}
\left(\frac{r_0}{5\,\text{au}}\right)^{3/2},
\end{equation}
which we note is a natural unit of mass for local shearing sheets.
In this work, we adopt $\rs = 0.8\RH$, where
\begin{equation}
\RH =
r_0\left(\frac{\Mp}{3 M_\star}\right)^{1/3} =
\Hg\left(\frac{\Mp}{3\Mth}\right)^{1/3}
\end{equation}
is the Hill radius and $\Hg \equiv \cs / \OmegaK$ is the vertical scale height of the gas.
Instead of using numerical differentiation, we analytically expand the gradient of equation~\eqref{E:gravpot} to compute the gravitational force exerted by the planet.

We initialise the system of gas and dust in equilibrium state as if the planet were not present.
The column density of the gas is initially constant at $\Sigmag = \Sigma_{\text{g},0}$, while the column density of the dust particles is initially constant at $\Sigmap = \Sigma_{\text{p},0} = Z\Sigma_{\text{g},0}$, where $Z$ is the solid abundance.
We randomly distribute the particles in the computational domain, and to reduce the Poisson noise in particle density, we use on average 36 particles per cell.
The initial velocities of the gas and the particles are given by the Nakagawa--Sekiya--Hayashi \citeyearpar[NSH;][]{NSH86} equilibrium solution \citep[see also][]{YJ16}.
We then gently increase the gravitational force of the planet from zero to actual magnitude over a duration of 10$P$, where $P \equiv 2\pi / \OmegaK$ is the local orbital period, by varying $G$ in equation~\eqref{E:gravpot}, as similarly done in \cite*{SYJ17}.

Because we consider the inviscid limit for this system, the density waves excited by the planet can freely propagate over long distances, and hence the boundary conditions require special attention.
To accommodate sufficient number of wavelengths, we adopt a particularly large computational domain, 32$\Hg$ in radial dimension and 64$\Hg$ in azimuthal dimension (which is equivalent to a disc aspect ratio of $\Hg / r_0 = \pi / 32 \simeq 0.1$).
After some experiments, we find that the most effective approach to establish the density waves without reflection is to use the standard sheared-periodic boundary conditions (\citealt{BN95}; \citealt*{HGB95}) while applying a damping zone near the radial boundary, as similarly done in global simulations \citep[e.g.,][]{VA07}.
Considering the force balance between the gas and the dust particles, we impose artificial source terms to the gas velocity $\ug$ and the particle velocity $\vp$ in the form of
\begin{align}
\tder{\ug}{t} &= \nu(x)\left(\tilde{\vec{u}}_\text{g} - \ug\right),\\
\tder{\vp}{t} &= \nu(x)\left(\tilde{\vec{v}}_\text{p} - \vp\right),
\end{align}
where the damping coefficient is defined by
\begin{align}
&\nu(x) \equiv \frac{1}{2}\left(10^{-2}P\right)^{-1}\times\nonumber\\
&\left\{
\begin{array}{ll}
1 + \cos\left[\pi\left(x - x_\text{min}\right) / \Hg\right],
   & \text{if }x_\text{min} \leq x < x_\text{min} + \Hg\\
0, & \text{if }x_\text{min} + \Hg \leq x \leq x_\text{max} - \Hg,\\
1 + \cos\left[\pi\left(x_\text{max} - x\right) / \Hg\right],
   & \text{if }x_\text{max} - \Hg < x \leq x_\text{max},
\end{array}
\right.
\end{align}
$\tilde{\vec{u}}_\text{g}$ and $\tilde{\vec{v}}_\text{p}$ are the NSH \citeyearpar{NSH86} equilibrium solutions for the gas and the particle velocities, respectively, and $\left[x_\text{min}, x_\text{max}\right]$ is the radial computational domain.

Finally, the small region near the planet also needs special treatment.
Numerous dust particles tend to accrete on to the planet and form a small cloud which is unresolved.
When the back reaction to the gas drag is on, the particles induce strong velocity shear in the gas around the planet, leading to numerical instability.
To compute the dynamics in this region properly, which is of interest in the study of pebble accretion \citep[see, e.g.,][]{JL17}, high resolution near the planet is required and hence is beyond the scope of this work.
Therefore, we include a region of radius 0.2$\RH$ around the planet such that any particle inside this region is removed.

In this work, we perform a systematic sweep of the parameter space in planet mass $\Mp$, dimensionless stopping time $\taus$, and solid abundance $Z$.
We consider $\Mp / \Mth = 0.1$, 0.3, and 1, $\taus = 10^{-3}, 10^{-2}, 10^{-1}$, and 1, and $Z \rightarrow 0$ (i.e., without back reaction), $Z = 0.1$, 0.3, and 1.
We fix the resolution at 16 cells per gas scale height $\Hg$ and conduct each simulation up to $t = 100P$.

\section{Limit of no dust back reaction} \label{S:nbr}

In this section, we summarise the results from the models without dust back reaction to the gas drag, i.e., in the limit of $Z \rightarrow 0$.
Without back reaction, the dust particles only passively respond to the gas dynamics, while the gaseous disc evolves as if the dust were not present.
This type of models has been studied relatively extensively in the literature, especially in global simulations.
Therefore, the presentation here serves as a consistency comparison with global models as well as a baseline to study the effects of the back reaction in the following sections.

The top row in Figs.~\ref{F:sigma1}--\ref{F:sigma3} shows the column densities of the gas and the dust particles side-by-side at the end of the simulations ($t = 100P$) for the models without back reaction.
The dimensionless stopping time $\taus$ increases from left to right along each row, while the planet mass $\Mp$ increases from Fig.~\ref{F:sigma1} to Fig.~\ref{F:sigma3}.
Given that in this case the dust particles do not exert any force on the gas, the evolution of the gas is independent of the properties of the dust present in the model, and thus only depends on the mass of the planet.

\begin{figure*}
	\includegraphics[width=\textwidth]{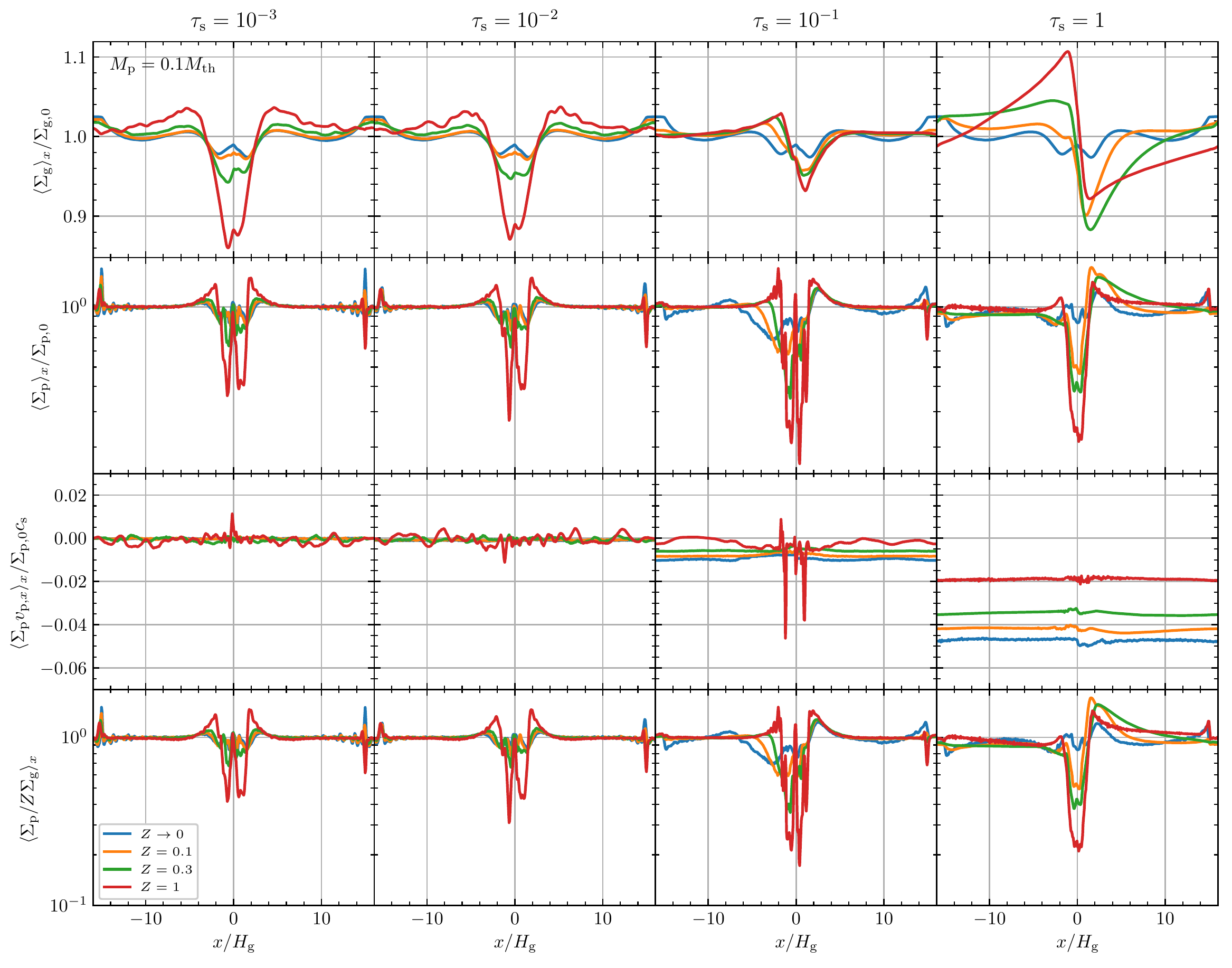}
    \caption{Azimuthally averaged profiles at the end of the simulations ($t = 100P$) from models with a planet of mass $\Mp = 0.1\Mth$, where $\Mth$ is the thermal mass (equation~\eqref{E:mth}).
        From top to bottom are the column density of the gas, the column density of the particles, the radial flux of the particles, and the solid-to-gas density ratio normalised by the solid abundance $Z$, respectively.
        Different line colours denote different $Z$s.
        The panels are organised such that the dimensionless stopping time $\taus$ increases from left to right.}
    \label{F:rprof1}
\end{figure*}

\begin{figure*}
	\includegraphics[width=\textwidth]{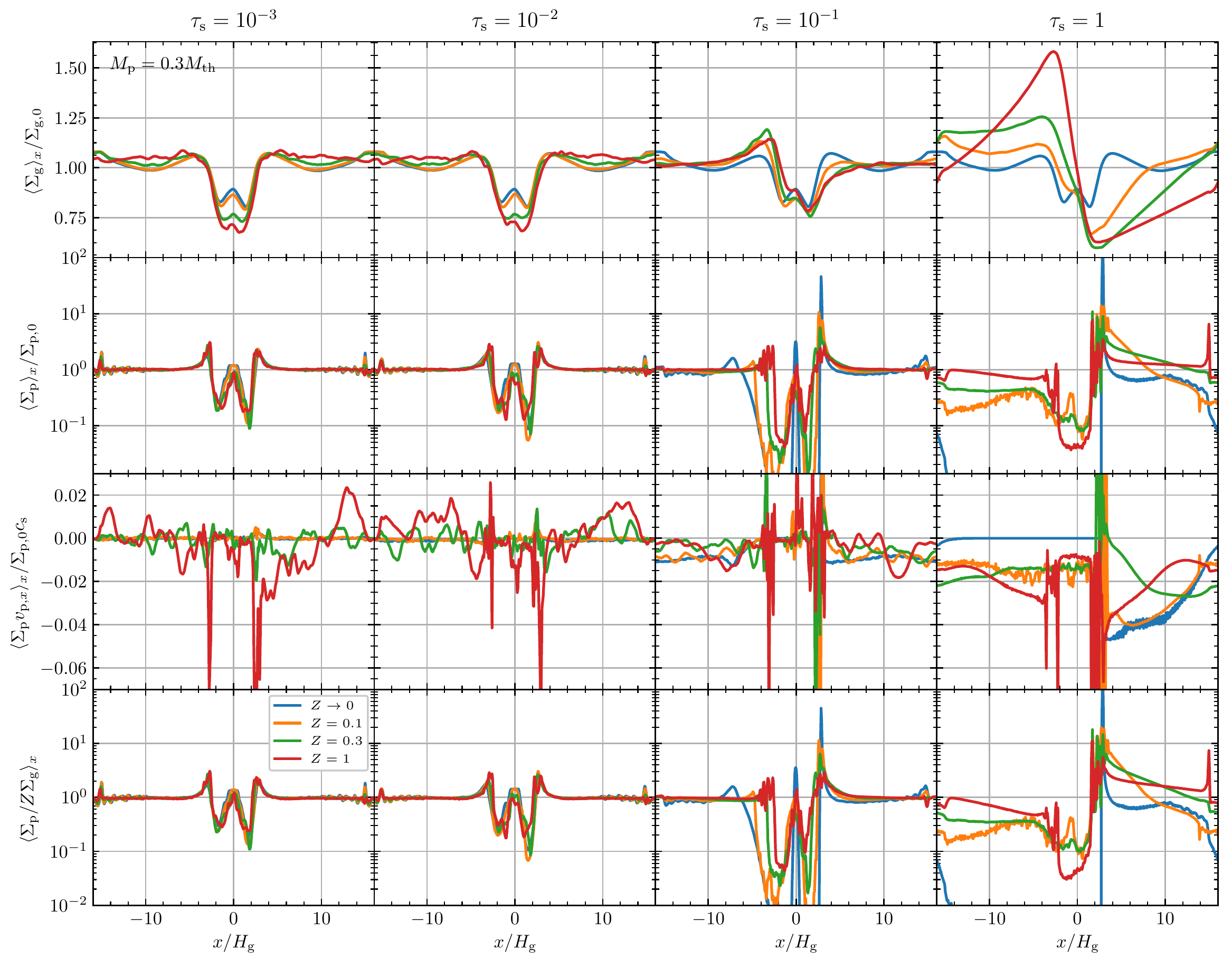}
    \caption{The same as Fig.~\ref{F:rprof1} except for models with $\Mp = 0.3\Mth$.}
    \label{F:rprof2}
\end{figure*}

\begin{figure*}
	\includegraphics[width=\textwidth]{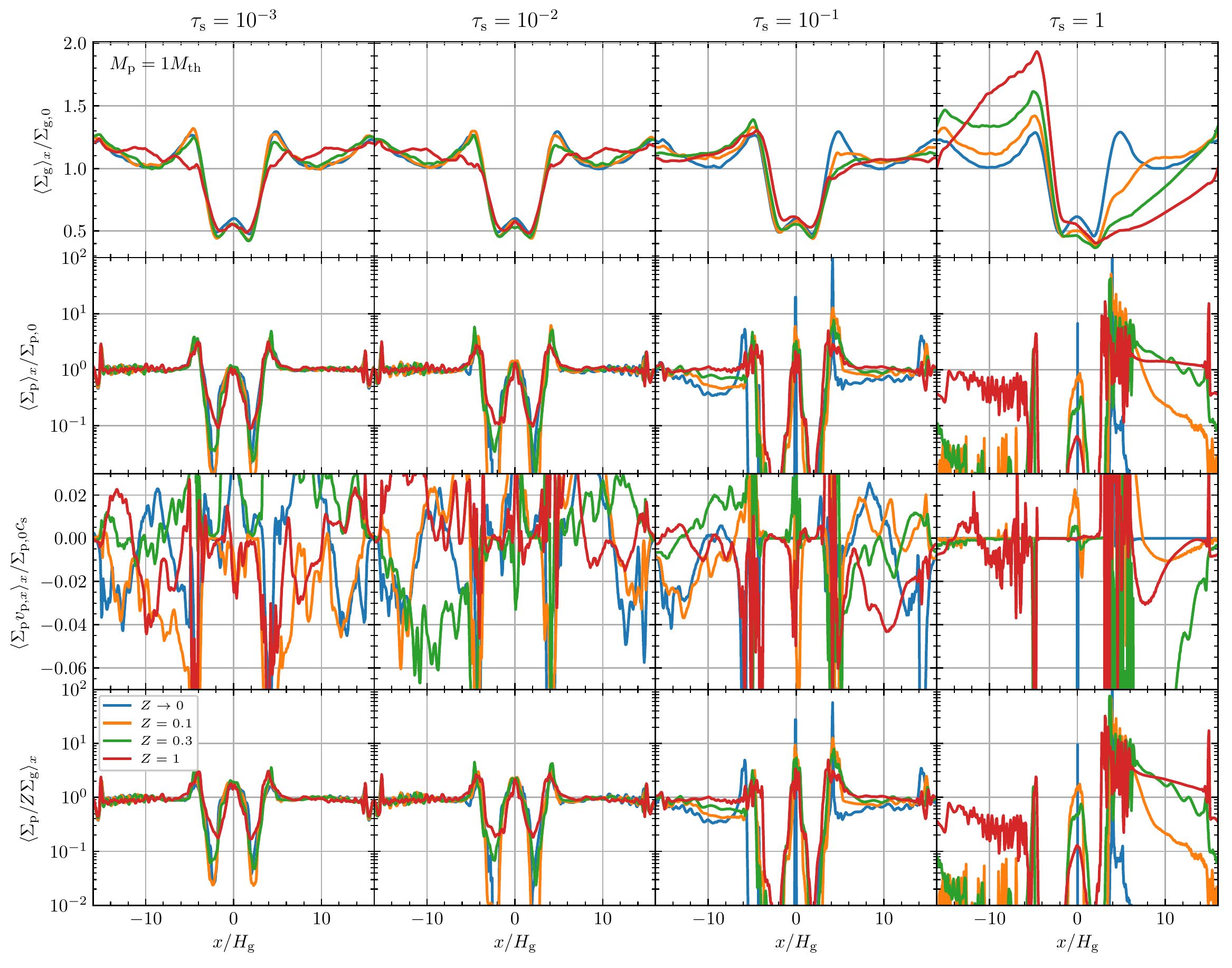}
    \caption{The same as Fig.~\ref{F:rprof1} except for models with $\Mp = 1\Mth$.}
    \label{F:rprof3}
\end{figure*}

As shown by the gas density on the left of each double panel, a planet can induce spiral arms, a gap, and/or vortices in the gaseous disc in the vicinity of its orbit.
Conspicuous two-arm spiral density wave in the gas is excited by the planet, wraps around the azimuthal periodic boundary, and propagates radially until being damped by the radial boundary.
When in the linear regime ($\Mp \ll \Mth$), the morphology of the wave is only a function of the Keplerian shear and the speed of sound, and is independent of the planet mass \citep{GT79}.
If the planet mass is non-negligible compared to the thermal mass, on the other hand, the spirals open up more as the planet mass increases \citep{ZD15,BZ18}.\footnote{It have been suggested that such a non-linear effect could explain the open spirals in recent observations with near-infrared scattered light \citep{DZ15}.}
In the meantime, the torque exerted by the planet drives the gas away from the orbit of the planet, and a gap begins to develop \citep{GT80}.
Given that the gas is inviscid, the depth of the gap continues to increase with time \citep*{DRS11}.
The blue lines in the top panels of Figs.~\ref{F:rprof1}--\ref{F:rprof3} show the azimuthal average of the gas density as a function of radial position at the end of the simulations.
The more massive the planet, the faster the gap becomes deeper.
The small bump near the planet orbit in the radial profile represents some of the gas being trapped in the horseshoe region (cf., Figs.~\ref{F:sigma1}--\ref{F:sigma3}).
When the gap is deep enough, as in the case of $\Mp = 1\Mth$ shown in Fig.~\ref{F:sigma3}, the gap edge is subject to the Rossby wave instability \citep{LL99}, and hence one large-scale vortex appears on each side of the gap \citep{VA07}.

When dust particles have a dimensionless stopping time of $\taus < 0.1$, they are tightly coupled to the gas.
Therefore, the morphology of these dust particles assimilates to that of the underlying gas, as shown by the particle density on the right of each double panel in the top row of Figs.~\ref{F:sigma1}--\ref{F:sigma3}.
The blue lines in the second-to-top panels of Figs.~\ref{F:rprof1}--\ref{F:rprof3} are the corresponding azimuthal average of the particle density as a function of radial position.
Similar to the gas, a gap in the dust distribution is formed.
The peak at the edge on each side of the gap, though, is slightly inside that of the gas.
Moreover, significant amount of dust particles remain in the horseshoe region.
When vortices are present, as in the case of $\Mp = 1\Mth$ shown in Fig.~\ref{F:sigma3}, the concentration of particles trapped in these vortices is also evident \citep[e.g.,][]{LJ09}, and the larger the stopping time $\taus$, the more concentrated the particles \citep{LL13}.

When the dimensionless stopping time $\taus \simeq 0.1$--1, the dust particles are marginally coupled to the gas, and the effects of their radial drift become apparent.
While accumulating near the outer edge of the gap in the gas, leading to a narrow dense dust ring, the particles continue to radially drift inward inside the orbit of the planet such that the gap in the particle distribution widens with time.
The larger the $\taus$, the faster this process of gap widening, which is a natural consequence of the faster radial drift as given by the NSH \citeyearpar{NSH86} equilibrium solution.
Moreover, the particles in the horseshoe region for the case of the planet mass $\Mp = 0.1\Mth$ have peculiar trajectories \citep{BP18}.
Particles radially drift faster than completing one orbit such that they form a smaller closed loop behind the planet.
When $\Mp \gtrsim 0.3\Mth$, the dust particles appear to be effectively depleted inside the orbit of the planet with time.

We note that because of the imposed background radial pressure gradient (Section~\ref{S:method}), a local pressure maximum does \emph{not} coincide with a local maximum in the gas density.
Instead, a local pressure maximum occurs where the gas density has a \emph{positive} radial gradient of $\partial\Sigmag / \partial x = +0.1(\Sigmag / \Hg)$.
The only favourable location in our model for this condition to occur is at the outer edge of the gap in the gas.
As shown by the blue lines in the top panels of Figs.~\ref{F:rprof1}--\ref{F:rprof3}, a true local pressure maximum exists near the outer edge and slightly inside the density maximum at the end of our simulations when the planet mass $\Mp \gtrsim 0.3\Mth$.
Indeed significant accumulation of dust particles occurs at this location, especially evident for marginally coupled ones ($0.1 \lesssim \taus \lesssim 1$).

The next-to-bottom panels in Figs.~\ref{F:rprof1}--\ref{F:rprof3} shows the azimuthally averaged radial flux of dust particles at the end of the simulations, in which we focus on the blue lines in this section.
For the models with a planet of mass $\Mp = 0.1\Mth$, the radial flux is close to be constant across radial position, indicating a state of quasi-steady equilibrium.
For the models with a planet of mass $\Mp \gtrsim 0.3\Mth$, on the other hand, the radial flux is halted near the outer edge of the gap driven by the planet, consistent with the location of a local pressure maximum.
This occurs irrespective of the dimensionless stopping time $\taus$, leaving the dust particles inside the planetary orbit gradually depleted with time as discussed above.

\section{Effects of dust back reaction}

When the back reaction to the gas drag by the dust particles is ignored, as summarised in Section~\ref{S:nbr}, the particles accumulate at the local pressure maximum located at the outer edge of the gap driven by the planet, or are trapped in a large-scale vortex.
In this case, the evolution of the system is invariant of the solid abundance $Z$.
As $Z$ becomes non-negligible, the local solid-to-gas density ratio can reach order unity or more at these dust traps (see the bottom panels of Figs.~\ref{F:rprof1}--\ref{F:rprof3}), indicating that the dust back reaction should become important in the dust-gas dynamics.
In this section, therefore, we activate the dust back reaction and consider its effects on the disc in comparison to the system without back reaction as presented in Section~\ref{S:nbr}.
As shown by Figs.~\ref{F:sigma1}--\ref{F:sigma3}, the dust back reaction indeed significantly changes the dynamics of the disc, when the initial solid-to-gas density ratio is as low as $Z \sim 0.1$.

\subsection{With tightly coupled dust particles (\texorpdfstring{$\taus < 0.1$}{t < 0.1})}
\label{SS:tcdp}

We first consider the models with tightly coupled dust particles, i.e., particles of dimensionless stopping time $\taus < 0.1$.
In this case, the relative velocity between each particle and its surrounding gas remains small and the particles closely follow the background gas flow.
Therefore, the system may be approximated by a single mixture with heavier inertia contributed by the dust and hence a smaller effective speed of sound \citep{LP14,LY17}.

\subsubsection{Spiral structure} \label{SSS:spiral1}

As discussed in Section~\ref{S:nbr}, the spiral structure depends on the planet mass $\Mp$ in the non-linear regime, i.e., when $\Mp$ is non-negligible as compared to the thermal mass $\Mth$.
Nevertheless, to simplify the complexity and study the effects of the dust back reaction on the spiral structure, we resort to the linear theory as a basis and use our models with the lowest mass $\Mp = 0.1\Mth$ for comparison.
According to the linear analysis of a gas-only disc, the openness of the spiral arms excited by the planet should scale with the local scale height of the gas $\Hg$ \citep{GR01,OL02,DR11}.
Extending this result to a dust-gas mixture implies that the higher the solid abundance $Z$, the more tightly wound the spiral arms should become because of the smaller effective scale height.
We test this hypothesis from our models in this section.

\begin{figure*}
\begin{center}
\subfigure[Tightly coupled particles\label{F:spiral1}]{
    \includegraphics[width=\columnwidth]{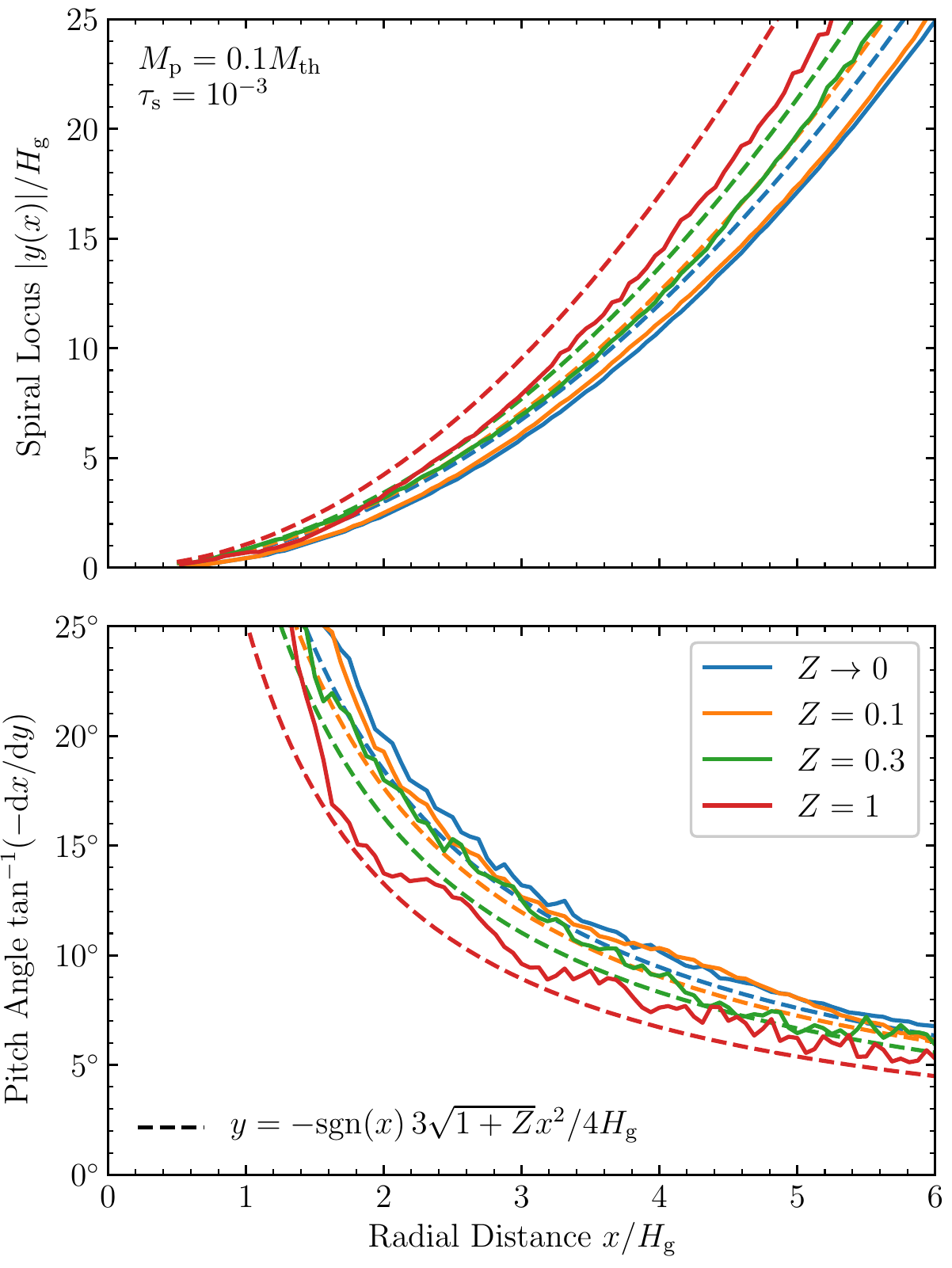}}
\subfigure[Marginally coupled particles\label{F:spiral2}]{
    \includegraphics[width=\columnwidth]{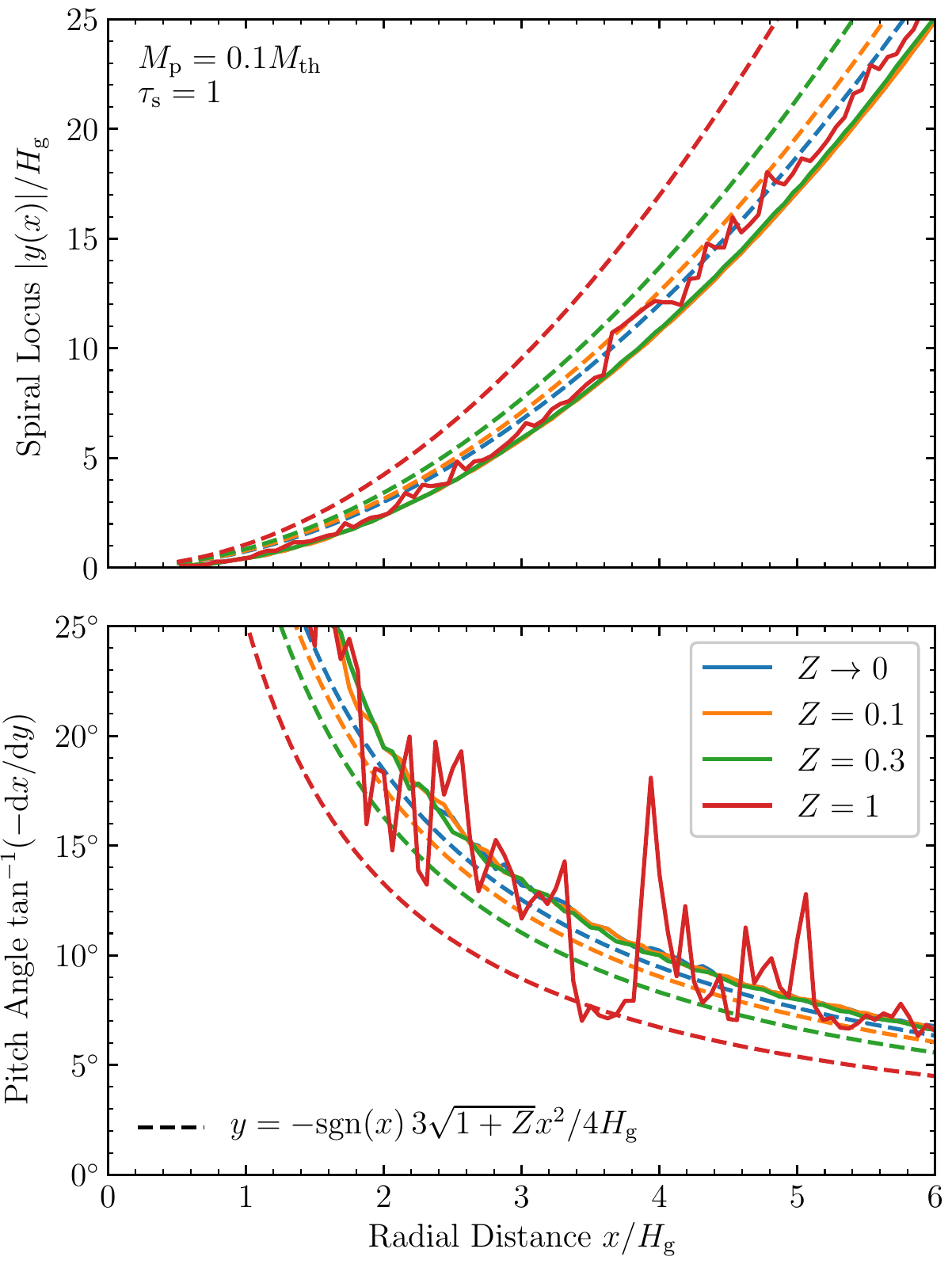}}
\caption{Locus (\textit{top}) and pitch angle (\textit{bottom}) of the spiral arm as a function of radial distance to the planet (\emph{solid} lines), measured from the models with a planet of mass 0.1$\Mth$ and dust particles of dimensionless stopping time (a)~$\taus = 10^{-3}$ and (b)~$\taus = 1$.
    Different colours represent different solid abundances $Z$.
    The \emph{dashed} lines are the corresponding theoretical asymptotic limit from the linear theory (equation~\eqref{E:spiral}).}
\label{F:spiral}
\end{center}
\end{figure*}

The solid lines in Fig.~\ref{F:spiral1} show the locus and the corresponding pitch angle of the spiral arm on the right side of the planet ($x > 0$) for models with a planet of mass $\Mp = 0.1\Mth$ and dust particles of dimensionless stopping time $\taus = 10^{-3}$.
We measure the locus of the spiral $y(x)$ by tracing the local maximum of the gas density along the spiral wake stemming from the planet.
We then use sixth-order centred differences to find the derivative $\mathrm{d}y / \mathrm{d}x$ and use it to compute the pitch angle $\tan^{-1}(-\mathrm{d}y / \mathrm{d}x)$.
Because this procedure introduces noise near the Nyquist frequency, we smooth the derivative $\mathrm{d}y / \mathrm{d}x$ by running averages of eight points.

For comparison, the dashed lines in Fig.~\ref{F:spiral1} are the asymptotic limit
\begin{equation} \label{E:spiral}
y \approx -\mathrm{sgn}(x)\frac{3 x^2}{4 \tilde{H}}
  = -\mathrm{sgn}(x)\sqrt{1 + Z}\frac{3 x^2}{4\Hg}
\end{equation}
for the corresponding models, where we have replaced the gas scale height $h$ in equation~(3) of \cite{DR11} with the effective scale height of a dust-gas mixture
\begin{equation}
\tilde{H} \equiv \frac{\tilde{c}_\text{s}}{\OmegaK} = \frac{\Hg}{\sqrt{1 + Z}},
\end{equation}
in which
\begin{equation} \label{E:ess}
\tilde{c}_\text{s} = \frac{\cs}{\sqrt{1 + Z}}
\end{equation}
is the effective speed of sound of the mixture \citep{LP14,CL18}.

Fig.~\ref{F:spiral1} indicates that the spiral arms indeed become less open with increasing solid abundance $Z$.
We note that the pitch angle is systematically larger than the corresponding asymptotic limits, which is due to the non-linear propagation of the density waves \citep{GR01} or the interference of their excitation \citep{OL02} with a planet of appreciable mass.
Nevertheless, the pitch angle of the spiral wakes in the models do approach to the corresponding asymptotes (derivative of equation~\eqref{E:spiral}).
Although it appears that the larger $Z$, the less rapidly the wake approaches to the limit, this comparison confirms qualitatively the expected effect of the back reaction by the dust on the openness of the spiral structure excited by a planet.
However, we show in Section~\ref{SSS:spiral2} that this property is not applicable for marginally coupled dust particles.

\subsubsection{Gap structure} \label{SSS:gap1}

We next turn our attention to the gap structure driven by the planet.
The top panels in Figs.~\ref{F:rprof1}--\ref{F:rprof3} indicate that with tightly coupled dust particles ($\taus < 0.1$), the higher the solid abundance $Z$, the deeper the gap in the gas becomes, except for the case with a planet of mass $\Mp = 1\Mth$, in which the gap depth is appreciably less sensitive to $Z$.
On the other hand, the gap width appears not to depend on the solid abundance.

\begin{figure}
\includegraphics[width=\columnwidth]{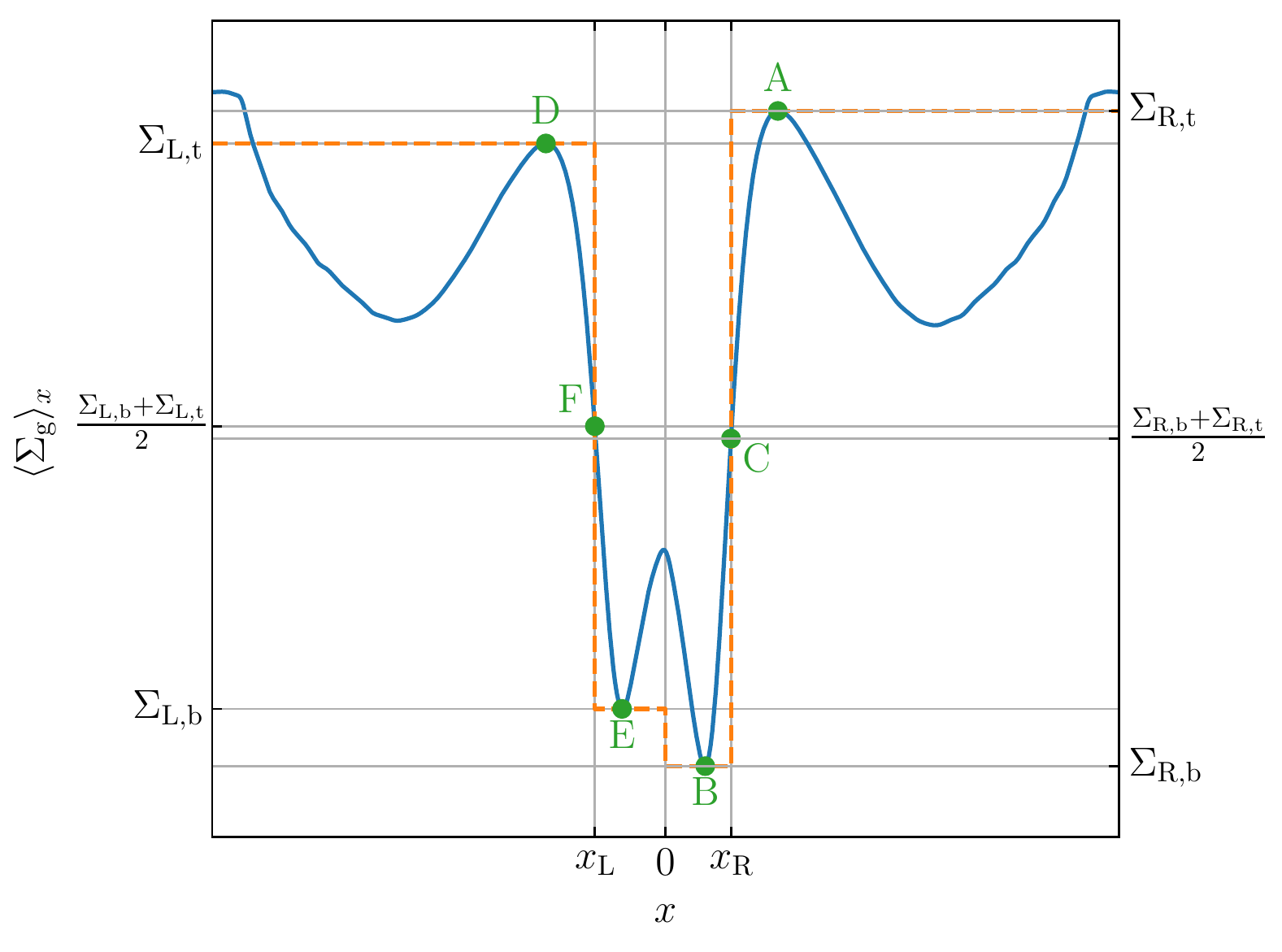}
\caption{Schematic diagram for measuring the depth and width of a gap opened by a planet.
    See Section~\ref{SSS:gap1} for the procedure.}
\label{F:gapsch}
\end{figure}

\begin{figure*}
\begin{center}
\subfigure[$\Mp = 0.1\Mth$\label{F:gap1}]{
    \includegraphics[width=0.95\columnwidth]{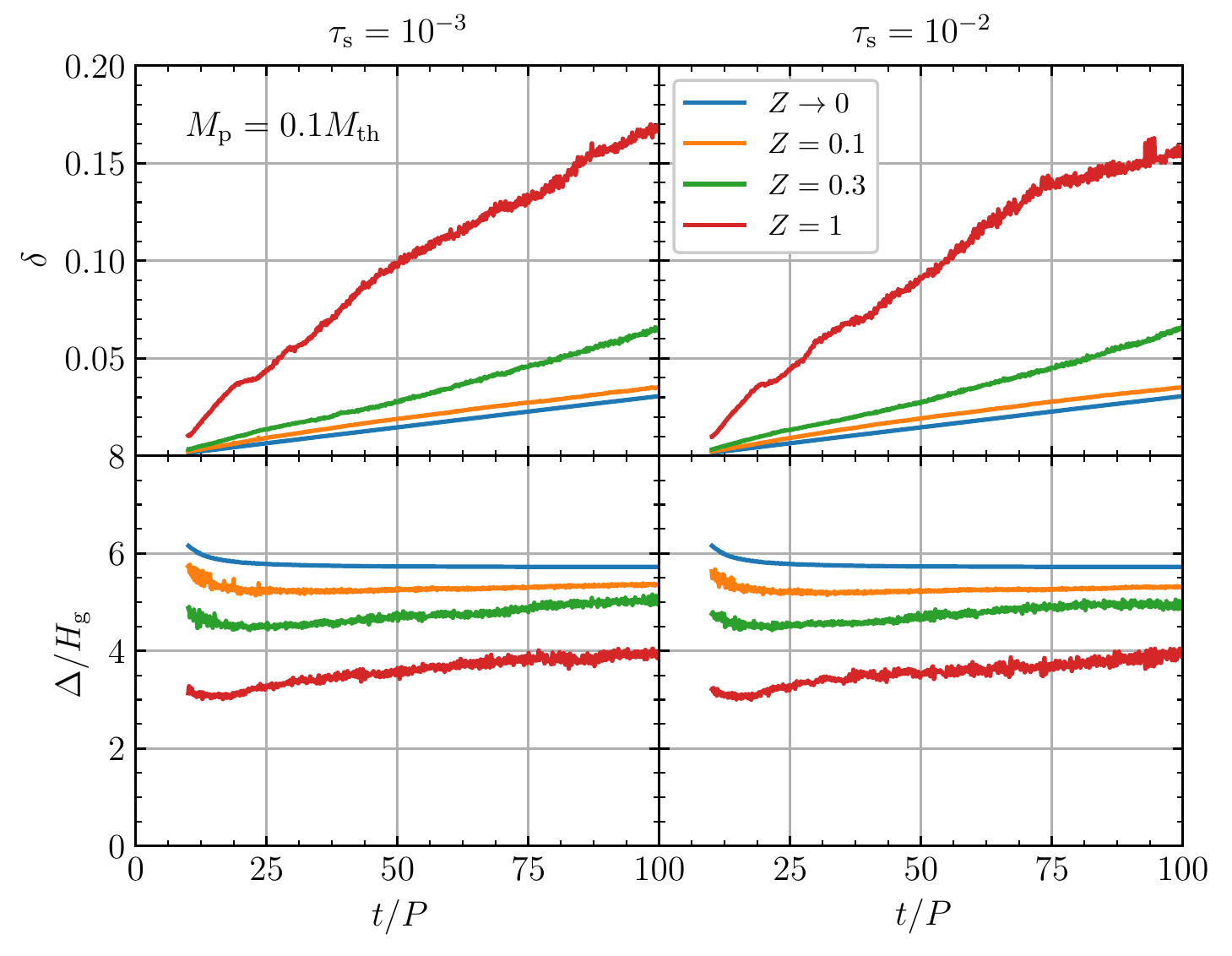}}\hspace{16pt}
\subfigure[$\Mp = 0.3\Mth$\label{F:gap2}]{
    \includegraphics[width=0.95\columnwidth]{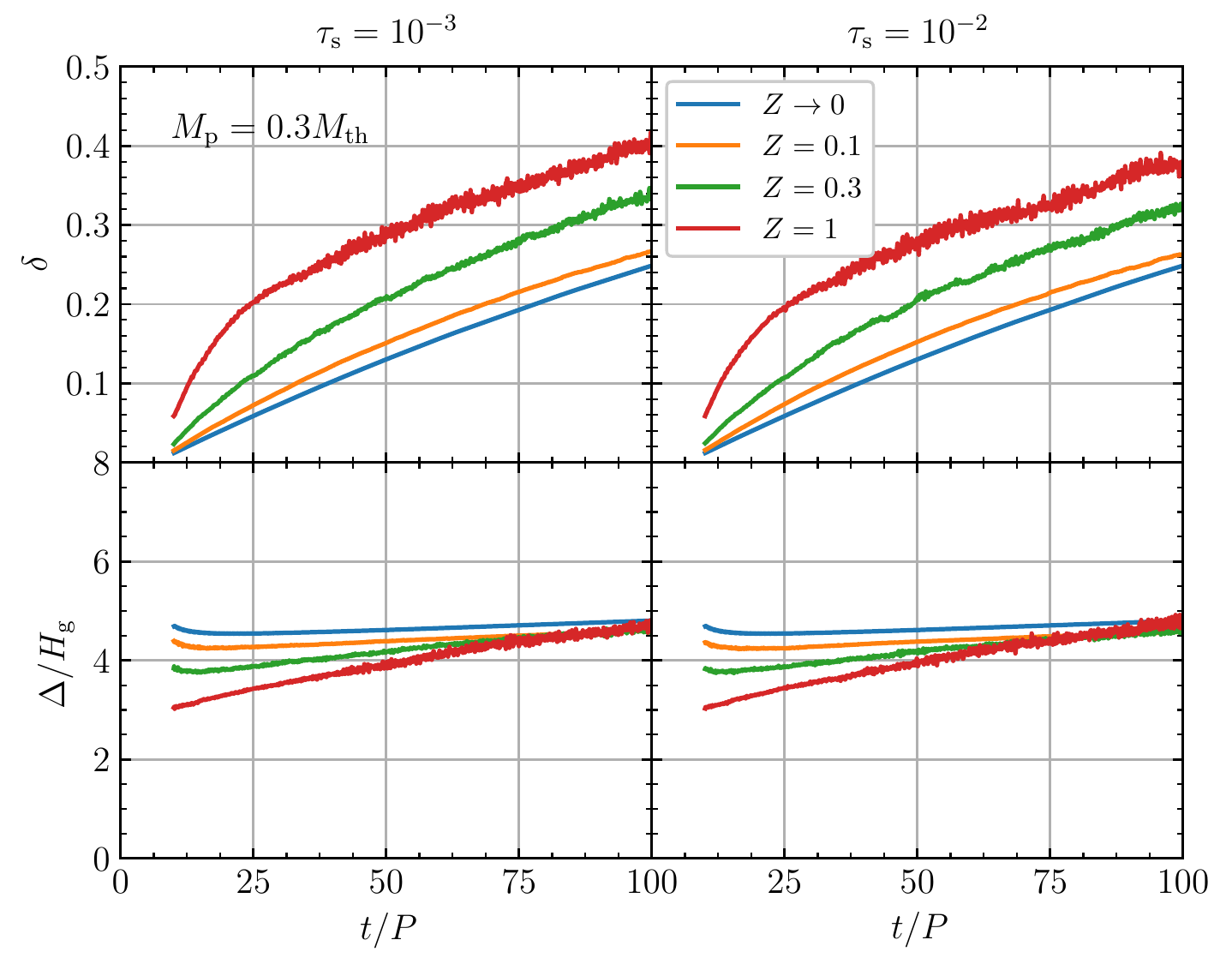}}\\
\subfigure[$\Mp = 1\Mth$\label{F:gap3}]{
    \includegraphics[width=1.5\columnwidth]{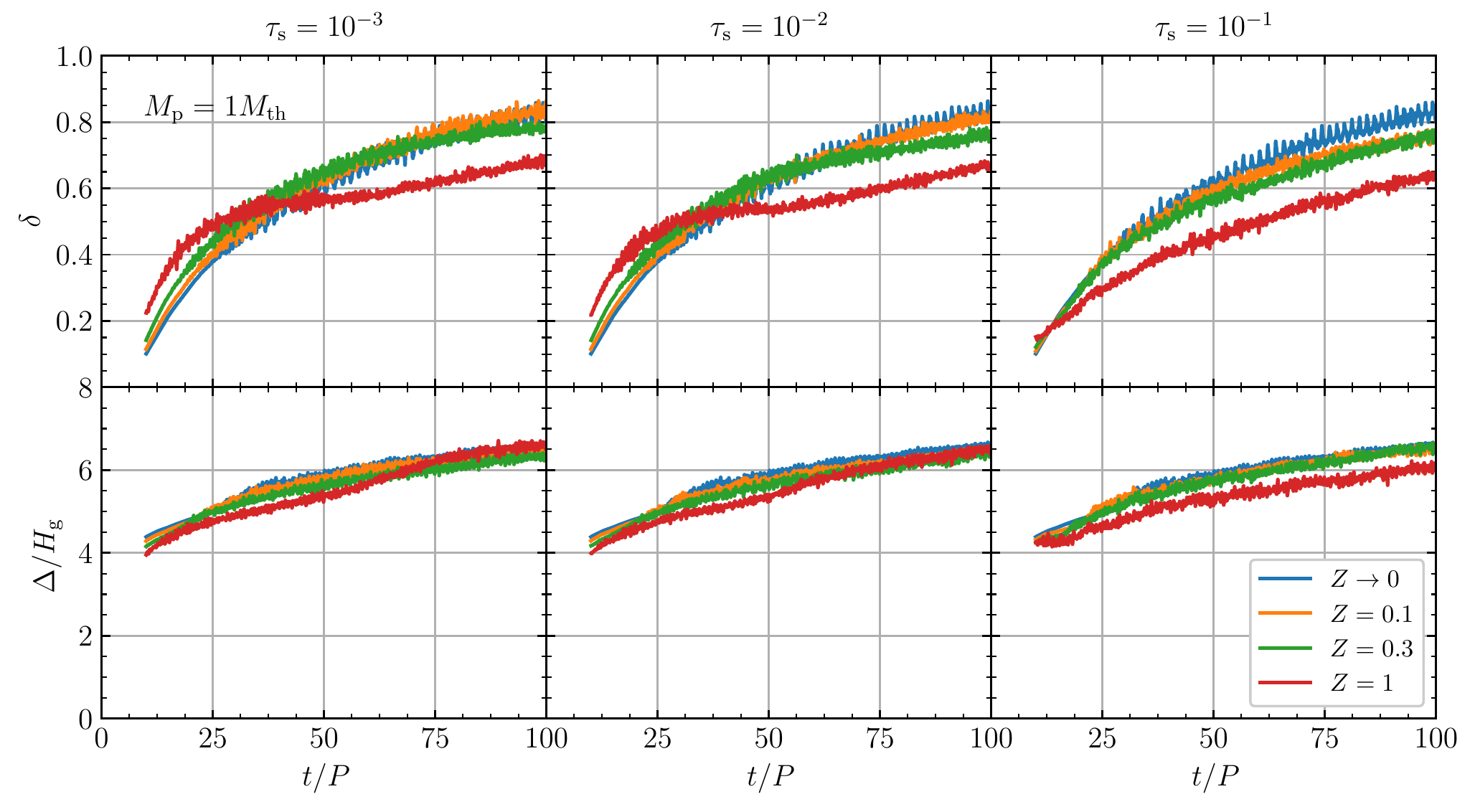}}
\caption{Evolution of the depth (\textit{top}) and width (\textit{bottom}) of the gap in the gas driven by a planet of mass (a)~$\Mp = 0.1\Mth$, (b)~$\Mp = 0.3\Mth$, and (c)~$\Mp = 1\Mth$.
    The dimensionless stopping time $\taus$ increases from the left column to the right, and lines of different colours represent different solid abundance $Z$.}
\label{F:gap}
\end{center}
\end{figure*}

To be more quantitative, we further measure the depth and width of the gap in the gas for each model as follows.
The solid line in Fig.~\ref{F:gapsch} shows a representative profile of azimuthally averaged gas density.
To detect the right edge of the gap, we first find the maximum of the local maxima on the right side of the planet (damping zone excluded), as represented by point~A, at which the density is defined by $\Sigma_\text{R,t}$.
Then we find the minimum of the local minima in between the planet and point~A, as represented by point~B, at which the density is defined by $\Sigma_\text{R,b}$.
We define the location of the right edge $x_\text{R}$ as the point where the profile crosses the midpoint between $\Sigma_\text{R,t}$ and $\Sigma_\text{R,b}$, as shown by point~C.
To detect the left edge of the gap $x_\text{L}$, we similarly find the points~D, E, and~F, and the corresponding densities $\Sigma_\text{L,t}$ and $\Sigma_\text{L,b}$.
Finally, we define the relative depth of the gap as
\begin{equation}
\delta \equiv
\frac{(\Sigma_\text{L,t} - \Sigma_\text{L,b}) + (\Sigma_\text{R,t} - \Sigma_\text{R,b})}
     {2\Sigma_\text{g,0}}
\end{equation}
and the width of the gap as
\begin{equation}
\Delta \equiv x_\text{R} - x_\text{L}.
\end{equation}
The resulting evolution of the depth $\delta$ and width $\Delta$ of the gap measured from all of our models are shown in Fig.~\ref{F:gap}.

The dependence of the depth evolution on our model parameters for tightly coupled dust particles ($\taus < 0.1$) can be summarised as follows.
The depth continuously increases with time in all of our models and approximately $\delta \propto \sqrt{t}$, which is consistent with our assumption of the inviscid limit \citep{DRS11}.
The more massive the planet, the more rapidly the gap grows deeper.
More importantly, it appears that the higher the solid abundance $Z$, the deeper the gap compared at the same instant of time.
The exception is for models with a planet of mass $\Mp = 1\Mth$, in which the depth of the gap is much less sensitive to the solid abundance, as shown by Figs.~\ref{F:sigma3}, \ref{F:rprof3}, and~\ref{F:gap3}.
We note that the measurement of the depth $\delta$ for the models with $\Mp = 1\Mth$ and $Z = 1$ is deceptively lower, because of the depreciation in the gas driven by the dust-gas vortices near the top of the edges (see Fig.~\ref{F:sigma3} and Section~\ref{SSS:vortex1}).

The trend of a deeper gap with increasing solid abundance $Z$ can be understood by the sensitive dependence of the torque $\Gamma$ exerted by the planet on the local scale height of the gas, i.e., $\Gamma \propto \Hg^{-2}$ (e.g., \citealt*{TTW02}; \citealt{PB10}).
Extending this result to a dust-gas mixture \citep{CL18}, $\Gamma \propto \tilde{H}^{-2} \propto (1 + Z)$ and hence the larger $Z$, the stronger the torque and the deeper the gap is.
We note that this relationship can be interpreted exactly as $\Gamma = \Gamma_\text{g} + \Gamma_\text{p} \propto \Sigmag + \Sigmap \approx \Sigmag (1 + Z)$, where $\Gamma_\text{g}$ and $\Gamma_\text{p}$ are the torques on the gas and the dust components, respectively.
On the other hand, the independence of the gap depth on $Z$ for the models with $\Mp = 1\Mth$ may be a consequence of the appreciably depleted dust content in the gap region compared to models with $\Mp \lesssim 0.3\Mth$, as shown by the azimuthally averaged solid-to-gas density ratio in the bottom panels of Figs.~\ref{F:rprof1}--\ref{F:rprof3}.
Therefore, it appears that the system is consistent with the approximation of a single dust-gas mixture, as in the case of the spiral structure discussed in Section~\ref{SSS:spiral1}.

\begin{figure}
\includegraphics[width=\columnwidth]{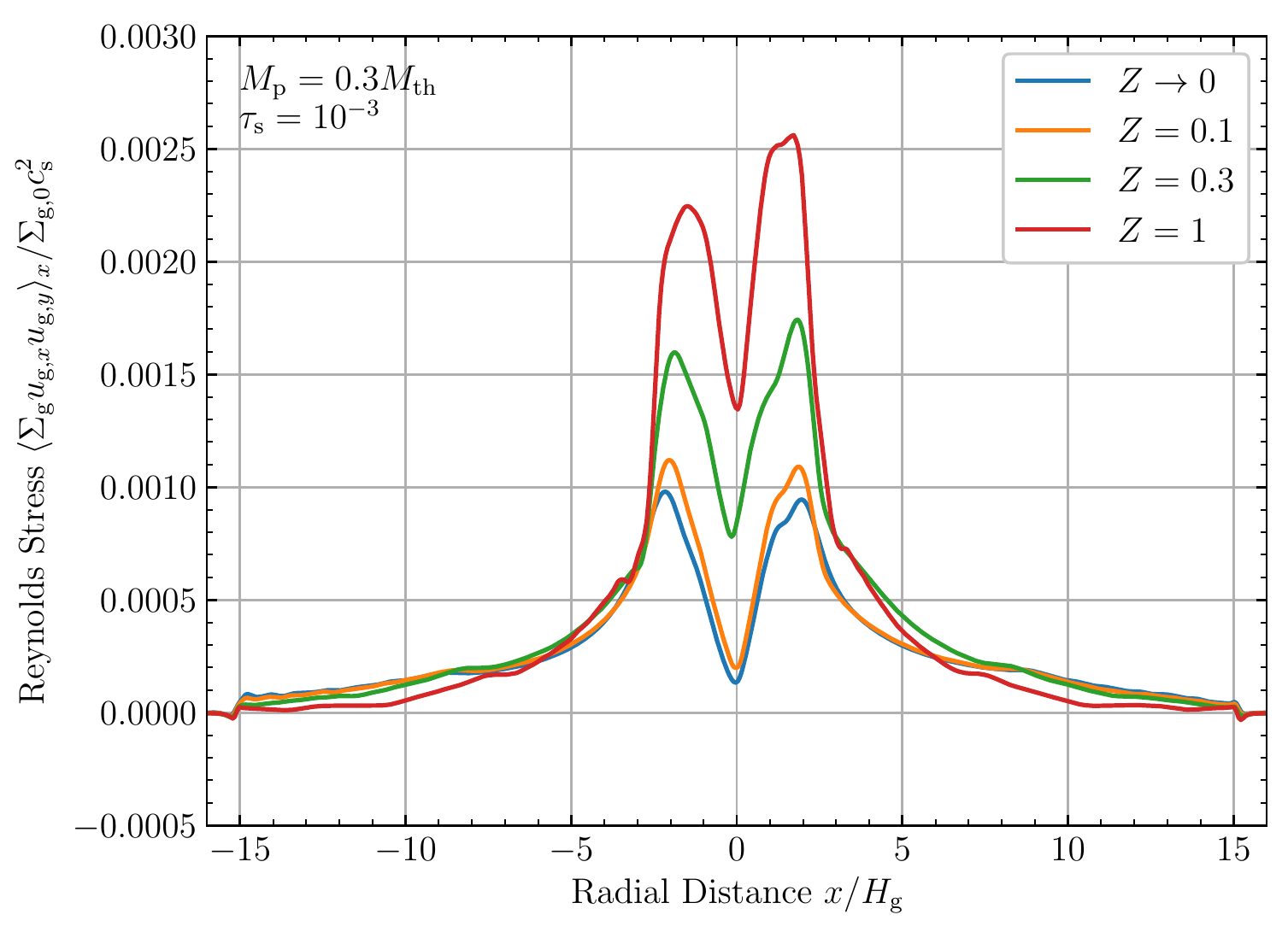}
\caption{Azimuthally averaged Reynolds stress as a function of radial distance for the models with a planet of mass $\Mp = 0.3\Mth$ and dust particles of dimensionless stopping time $\taus = 10^{-3}$.
    Different colours represent different solid abundances $Z$.
    The profiles are time-averaged over the last 20$P$ of the simulations.}
\label{F:reynolds}
\end{figure}

To be more quantitative, we plot in Fig.~\ref{F:reynolds} the time and azimuthally averaged Reynolds stress -- which is equivalent to the angular momentum flux -- as a function of radial distance to the planet from the models with planet mass $\Mp = 0.3\Mth$ and dimensionless stopping time $\taus = 10^{-3}$.
The peak of the Reynolds stress correlates with the strength of the torque density while the angular momentum is deposited where the stress begins to decline \citep{DR11}.
From Fig.~\ref{F:reynolds}, the maximum stress is larger by a factor of 1.1, 1.8, and 2.6  when the solid abundance $Z = 0.1, 0.3$, and 1 with respect to the case of $Z \rightarrow 0$.
In comparison, the predicted torque ratios are 1.1, 1.3, and 2, respectively.
The apparently larger stress than predicted is likely contributed by the dust-gas vortices induced by the dust back reaction (Section~\ref{SSS:vortex1}).
A similar consideration of additional contribution to angular momentum flux may be found in a system with MHD turbulence \citep*{ZSR13}.
Finally, we note that the disparity is even larger for the models with $\Mp = 0.1\Mth$, indicating stronger contributions of the dust-gas vortices to the angular momentum flux.

We next observe the evolution of the gap width for tightly coupled dust particles ($\taus < 0.1$).
From the top panels of Fig.~\ref{F:rprof1}, we see that any gap with a depth contrast of $\delta \lesssim 10\%$ appears not well developed, and the measurement of its width $\Delta$ becomes sensitive to the exact density profile of each edge \citep[see also][]{ZZ18}.
For the models with a planet of mass $\Mp = 0.1\Mth$, the gap has not well developed within 100$P$ when the solid abundance $Z \lesssim 0.3$ (Fig.~\ref{F:gap1}), while the gap becomes well developed after an initial period for the models with $\Mp = 0.3\Mth$ (Fig.~\ref{F:gap2}).
Taking this into account, we see from the bottom panels of Fig.~\ref{F:gap} that the gap width also in general increases with time.
Moreover, the gap is initially narrower for larger $Z$.
However, for the same planet mass but irrespective of $Z$, the gap appears to approach to similar width given enough time.
This indicates that the effective radial distance for the planet to clear the materials near its orbit is not sensitive to the solid abundance.

Finally, the second-to-top panels in Figs.~\ref{F:rprof1}--\ref{F:rprof3} shows the azimuthally averaged column density of the dust particles.
With tightly coupled particles ($\taus < 0.1$), the gap structure driven  by the planet is more prominent in the dust component than in the gas.
The dust accumulates at the pressure maximum outside of the planet while simply being pushed away inside of the planet (Section~\ref{S:nbr}).
The dust particles in between are significantly depleted except in the horseshoe region.
The more massive the planet, the more depleted the dust component, while most of the dust materials in the horseshoe region remains irrespective of the planet mass.

More interesting is the dependence of the dust gap on the solid abundance $Z$.
For the models with a planet of mass $\Mp = 0.1\Mth$, the higher $Z$, the more prominent is the dust gap (see also Fig.~\ref{F:sigma1}).
This behaviour simply follows how well the gas gap is developed, as discussed above.
For $\Mp \gtrsim 0.3\Mth$, on the other hand, the higher $Z$, the \emph{less} prominent is the dust gap.
This appears to be closely related with the instability generated near the edge of the dust gap, as discussed in the following section (Section~\ref{SSS:vortex1}).

\subsubsection{Vortices} \label{SSS:vortex1}

When the back reaction from the dust particles to the gas drag is activated, several new phenomena appear.
First, in the very vicinity of the planet, small dust-gas vortices are generated randomly in time when dust particles attempt to make a U-turn inside the Hill radius of the planet (Figs.~\ref{F:sigma1}--\ref{F:sigma3}).
These vortices continue to follow the horseshoe trajectories near the co-orbital radius without any further change.
Moreover, the higher the solid abundance $Z$, the stronger this effect becomes.
As discussed in Section~\ref{S:method}, the trajectories inside the Hill radius of the planet is not well resolved, and hence to confirm the validity of and further study this effect requires future investigations.

Second, the dust back reaction indeed tends to make the edge of the dust gap unstable, especially when sufficient dust material is accumulated near the edge \citep[see also][]{PLR19}.
As shown in Fig.~\ref{F:sigma2}, the instability occurs within 100$P$ for the models with tightly coupled particles ($\taus < 0.1$) and a planet of mass $\Mp = 0.3\Mth$, when the solid abundance $Z \gtrsim 0.3$.
This instability is of Kelvin--Helmholz type because of a steep gradient in dust-to-gas density ratio, as shown by the bottom panels of Fig.~\ref{F:rprof2}, which drives a gradient in gas velocity due to the dust back reaction (the higher the density ratio, the closer to be Keplerian the gas is accelerated).
In other words, a dust ring should have a minimum width below which the velocity shear becomes too strong and the system is unstable \citep[e.g.,][]{YM10}.
When this occurs, the axisymmetric narrow ring of dust at the edge is broken into several small-scale vortices.
Moreover, the higher $Z$, the more such vortices are generated.
Interestingly, this behaviour tends to regulate the dust accumulation at the gap edge in  such a way that $\Sigmap / Z\Sigmag \simeq 2$, as shown in Fig.~\ref{F:rprof2}.

Finally, as discussed in Section~\ref{S:nbr}, a planet of mass $\Mp = 1\Mth$ is sufficiently heavy to drive large-scale vortices near the edge of the gas gap, and dust particles are trapped in and coevolve with these vortices.
However, when the dust back reaction becomes important, the large-scale vortices are broken into numerous small-scale vortices (see also \citealt{FL14}; \citealt*{CZS15}; \citealt*{RKL15}; \citealt{SML16}; \citealt{SM19}).
As shown in Fig.~\ref{F:sigma3}, this occurs when the solid abundance $Z \gtrsim 0.3$ for tightly coupled dust particles ($\taus < 0.1$).
(A similar behaviour also occurs for marginally coupled dust particles; see Section~\ref{SSS:vortex2}.)
Moreover, the dust accumulation at the gap edge is similarly maintained at $\Sigmap / Z\Sigmag \simeq 2$ as in the case of smaller planets described above (Fig.~\ref{F:rprof3}).

\subsection{With marginally coupled dust particles (\texorpdfstring{$0.1 \lesssim \taus \lesssim 1$}{0.1 < t < 1})} \label{SS:mcdp}

We next consider the models with marginally coupled dust particles, i.e., particles which have a dimensionless stopping time of $0.1 \lesssim \taus \lesssim 1$.
In this case, the particles may not closely follow the background flow of the gas, and the dust-gas dynamics becomes deviated from the description of a single mixture as appropriate for tightly coupled dust particles considered in Section~\ref{SS:tcdp}.
We discuss such differences in the following sections.

\subsubsection{Spiral structure} \label{SSS:spiral2}

In Section~\ref{SSS:spiral1}, we demonstrate that the spiral structure becomes less open with increasing solid abundance $Z$ when the disc contains tightly coupled dust particles, as shown in Fig.~\ref{F:spiral1}.
However, this behaviour is not applicable to marginally coupled dust particles.
As shown in Fig.~\ref{F:spiral2}, the spiral locus and the corresponding pitch angle as a function of the radial distance to the planet does not change with $Z$ for the case of $\taus = 1$ (the variation in the case of $Z = 1$ is due to the difficulty with the presence of numerous small-scale dust-gas vortices).
Irrespective of $Z$, the pitch angle is close to the asymptotic limit (equation~\eqref{E:spiral}) as if the dust were not present (i.e., $Z \rightarrow 0$).
We note that, as shown by Fig.~\ref{F:rprof1}, the azimuthally averaged column density ratio remains similar to the initial value for $|x| \gtrsim 2\Hg$, i.e., $\langle\Sigmap / \Sigmag\rangle_x \simeq Z$.
Therefore, the effective speed of sound does not appear to be modified by the dust back reaction as in equation~\eqref{E:ess} when the dust particles are marginally coupled, and hence caution needs to be exercised when considering the system as a single mixture in this case.

\subsubsection{Gap structure} \label{SSS:gap2}

It has long been accepted that a planet can drive near its orbit a gap in the gas component of the disc.
However, appreciably different structure appears when the disc contains significant amount of marginally coupled dust particles ($0.1 \lesssim \taus \lesssim 1$).
As shown in the top panels of Figs.~\ref{F:rprof1}--\ref{F:rprof3}, the gas on both sides of the planetary orbit is pushed radially outwards.
Moreover, the higher the solid abundance $Z$, the more drastic the effect is.
As a result, the relatively radially symmetric gap structure about the planet in the limit of no dust back reaction becomes more and more skewed as $Z$ increases \citep[see also][]{CL19}.
A gap cannot even be well defined for $Z$ being as low as about 0.1 (see also Figs.~\ref{F:sigma1}--\ref{F:sigma3}) in almost all of our models.
The exceptions are the models with a planet of mass $\Mp = 1\Mth$, in which the gap structure remains significant when $\taus = 0.1$ and marginal when $\taus = 1$ and $Z = 0.1$.
Therefore, it appears that the smaller the planet mass $\Mp$ and the larger the dimensionless stopping time $\taus$ (but close to unity), the less conspicuous the gap structure in the gas is.

This behaviour could simply be understood by angular momentum conservation.
As the dust particles radially drift inwards toward the star due to the loss of their angular momentum to the head wind, the gas gains equal amount of angular momentum and hence radially moves outwards \citep[see also][]{GLM17}.
The strength of this effect can be seen via the NSH \citeyearpar{NSH86} equilibrium solution for the radial velocities of the gas and the dust particles
\begin{align}
\tilde{u}_{\text{g},x} &=
\frac{2\epsilon\taus}{(1 + \epsilon)^2 + \taus^2}\Delta u,\label{E:nshg}\\
\tilde{v}_{\text{p},x} &=
-\frac{2\taus}{(1 + \epsilon)^2 + \taus^2}\Delta u,\label{E:nshp}
\end{align}
respectively, where $\epsilon$ is the local dust-to-gas density ratio and $\Delta u$ is the velocity reduction in the gas defined in Section~\ref{S:method} \citep[see also][]{YJ16}.
For any given dimensionless stopping time $\taus$, the gas reaches its maximum outward radial drift when $\epsilon = \sqrt{\taus^2 + 1}$, at which
\begin{equation} \label{E:umax}
\left.\tilde{u}_{\text{g},x}\right|_\text{max} = \frac{\taus}{1 + \sqrt{\taus^2 + 1}}\Delta u.
\end{equation}
We note that equation~\eqref{E:umax} is a monotonically increasing function of $\taus$, asymptotically approaching $\Delta u$.
This maximum drift equals half of $\Delta u$ when $\taus = 4 / 3$.
Therefore, the modification of the gap structure in the gas due to this effect is the most noticeable when $\taus \sim \epsilon \sim 1$, which is consistent with the findings described above.
Finally, the modified structure should reach an equilibrium state by balancing its induced pressure gradient, as shown by the profiles in Figs.~\ref{F:rprof1}--\ref{F:rprof3}.

Equation~\eqref{E:umax} indicates that the time-scale for the dust back reaction to modify its background gas is similar to that of dust radial-drift as if $\epsilon \rightarrow 0$.
For the dimensionless stopping time $\taus \ll 1$, $\left.\tilde{u}_{\text{g},x}\right|_\text{max} \simeq \taus \Delta u / 2$, while for $\taus \gtrsim 1$, $\left.\tilde{u}_{\text{g},x}\right|_\text{max} \sim \Delta u$.
The crossing time for a structure of width $\Hg$ is then about $6P / \taus$ and $3P$, respectively.
As expected, we see this effect of the back reaction from marginally coupled  dust particles ($0.1 \lesssim \taus \lesssim 1$) within 100$P$, while not from tightly coupled dust particles ($\taus < 0.1$).

The dynamics of the marginally coupled dust particles is more complicated and so are the resulting morphological signatures.
Within 100$P$ for all the models with a planet of mass $\Mp = 0.1\Mth$, as shown in the top panels of Fig.~\ref{F:rprof1}, the gap in the gas is not developed deep enough and hence no true local pressure maximum is established radially outside of the planet (see Section~\ref{S:nbr}).
Therefore, marginally coupled (as well as tightly coupled) dust particles continue to radially drift through the gap.
As shown by the next-to-bottom panels of Fig.~\ref{F:rprof1}, the radial flux of the particles is roughly constant across all radial distance, and hence an equilibrium state in the dust component has been reached and maintained.
Moreover, a dust gap is established and it appears that the higher the solid abundance $Z$, the deeper the gap becomes (see also Fig.~\ref{F:sigma1}).
This phenomenon is related with the skewed profile in the gas component discussed above.
As the negative density gradient of the gas inside the planetary orbit becomes steeper with increasing $Z$, the dust particles in this region experience stronger head wind (larger $\Delta u$) and hence radially drift inward faster (equation~\eqref{E:nshp}), which results in a deeper dust gap.

When the planet has a mass of $\Mp \gtrsim 0.3\Mth$ and the dust back reaction is negligible ($Z \rightarrow 0$), a true local pressure maximum outside of the planetary orbit is established within 100$P$, where radial flux of dust particles is halted (Section~\ref{S:nbr}).
However, this pressure maximum is significantly weakened or even disappears as the solid abundance $Z$ increases, as shown in the top panels of Figs.~\ref{F:rprof2} and~\ref{F:rprof3}.
For the models with $\Mp = 0.3\Mth$, marginally coupled dust particles can again penetrate through the gap, when $Z \gtrsim 0.3$ and $Z \gtrsim 0.1$ for $\taus = 0.1$ and $\taus = 1$, respectively (Fig.~\ref{F:rprof2}).
When this occurs, the dust gap is well maintained, which is in drastic contrast to the case of $Z \rightarrow 0$ in which the dust is depleted inside of the planetary orbit (see also Fig.~\ref{F:sigma2}).
For the models with $\Mp = 1\Mth$, on the other hand, the pressure maximum remains present except for $\taus = 1$ and $Z = 1$ (Fig.~\ref{F:rprof3}).
In all cases (including $\taus = 1$ and $Z = 1$), however, the dust particles are well depleted near the planetary orbit except that some remains in the horseshoe region (see also Fig.~\ref{F:sigma3}).
Therefore, it appears that when $\Mp \gtrsim 1\Mth$, the gravity of the planet is strong enough to deplete its nearby orbits and impede the dust radial drift regardless of the presence of a pressure maximum.

Finally, for all cases where the dust cannot radially drift past the planetary orbit, the dust component inside of the orbit become depleted with time.
However, we note that the rate of depletion is slower with increasing solid abundance $Z$, as shown by the radial location of the dust front in Figs.~\ref{F:rprof2} and~\ref{F:rprof3} (see also Figs.~\ref{F:sigma2} and~\ref{F:sigma3}).
This phenomenon is predominantly due to the effect that the larger $Z$, the slower the dust radial drift, as depicted by equation~\eqref{E:nshp}.

\subsubsection{Vortices} \label{SSS:vortex2}

For the models with a planet of mass $\Mp = 0.1\Mth$ and marginally coupled dust particles ($0.1 \lesssim \taus \lesssim 1$), the only vortices within the duration of the simulations are generated near the planet, a similar effect observed in the models with tightly coupled dust particles.
See Section~\ref{SSS:vortex1} for its discussion.

Also similar to the models with tightly coupled dust particles discussed in Section~\ref{SSS:vortex1}, the narrow dust ring near the outer edge of the gap driven by a planet of mass $\Mp = 0.3\Mth$ becomes unstable with marginally coupled dust particles and numerous dust-gas vortices are generated.
However, the instability occurs \emph{earlier} than the models with tightly coupled particles and can be observed within the duration of the simulations for solid abundance as low as $Z \sim 0.1$ (Fig.~\ref{F:sigma2}).
This is due to the faster radial drift of the marginally coupled particles and hence faster accumulation in the dust ring.
On the other hand, the peak dust density near the edge at the end of the simulations appears to be maintained at $\Sigmap \sim 0.2\Sigma_{\text{g},0}$ for marginally coupled particles, as shown in Fig.~\ref{F:rprof2}.

In contrast to the region outside of the planetary orbit, the dust-gas vortices generated inside are significantly less with marginally coupled dust particles (Fig.~\ref{F:sigma2}).
This phenomenon can be understood by noting that the dust particles are relatively depleted inside the planetary orbit due to faster radial drift, reducing the solid-to-gas density ratio, as shown in Fig.~\ref{F:rprof2}.
The effect is especially evident for the models with the dimensionless stopping time of $\taus = 1$.

Finally, for the models with a planet of mass $\Mp = 1\Mth$ and marginally coupled dust particles, the large-scale lopsided vortex on either side of the planetary orbit is broken into numerous small-scale dust-gas vortices when the dust back reaction becomes important, as shown in Fig.~\ref{F:sigma3}.
This phenomenon is similar to the models with tightly coupled dust particles, as discussed in Section~\ref{SSS:vortex1}.
Interestingly, the large-scale lopsided vortex \emph{inside} the planetary orbit \emph{reappears after} the dust particles become depleted in the region because of their radial drift.
This occurs in the models with the dimensionless stopping time of $\taus = 1$ and the solid abundance $Z \lesssim 0.3$ (Fig.~\ref{F:sigma3}).

\section{Implications for Observations} \label{S:obs}

\begin{figure*}
\begin{center}
\subfigure[Low-mass planet ($\Mp < \Mth$)\label{F:summary1}]{
    \includegraphics[width=0.95\textwidth]{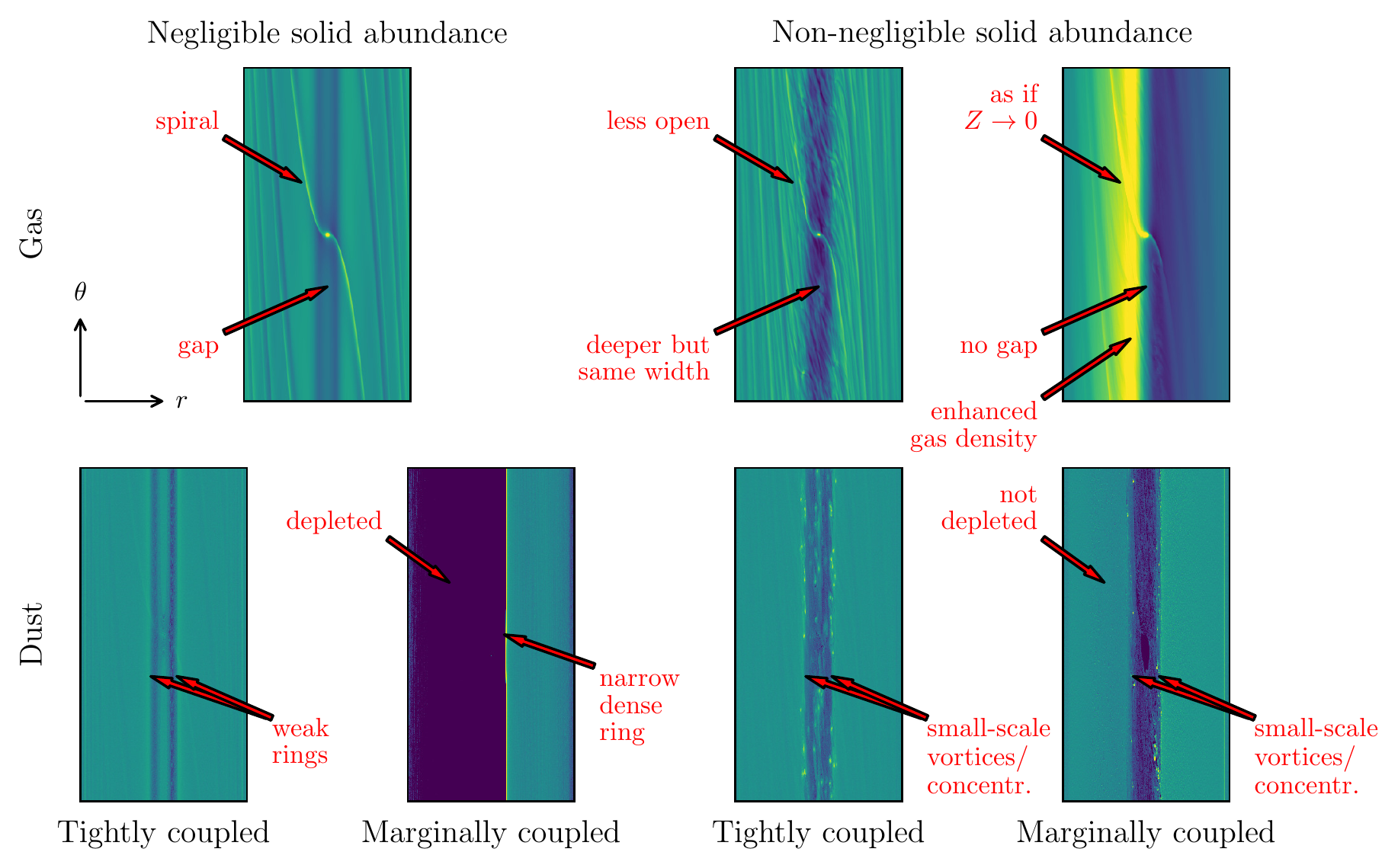}}\\
\subfigure[High-mass planet ($\Mp \gtrsim \Mth$)\label{F:summary2}]{
    \includegraphics[width=0.95\textwidth]{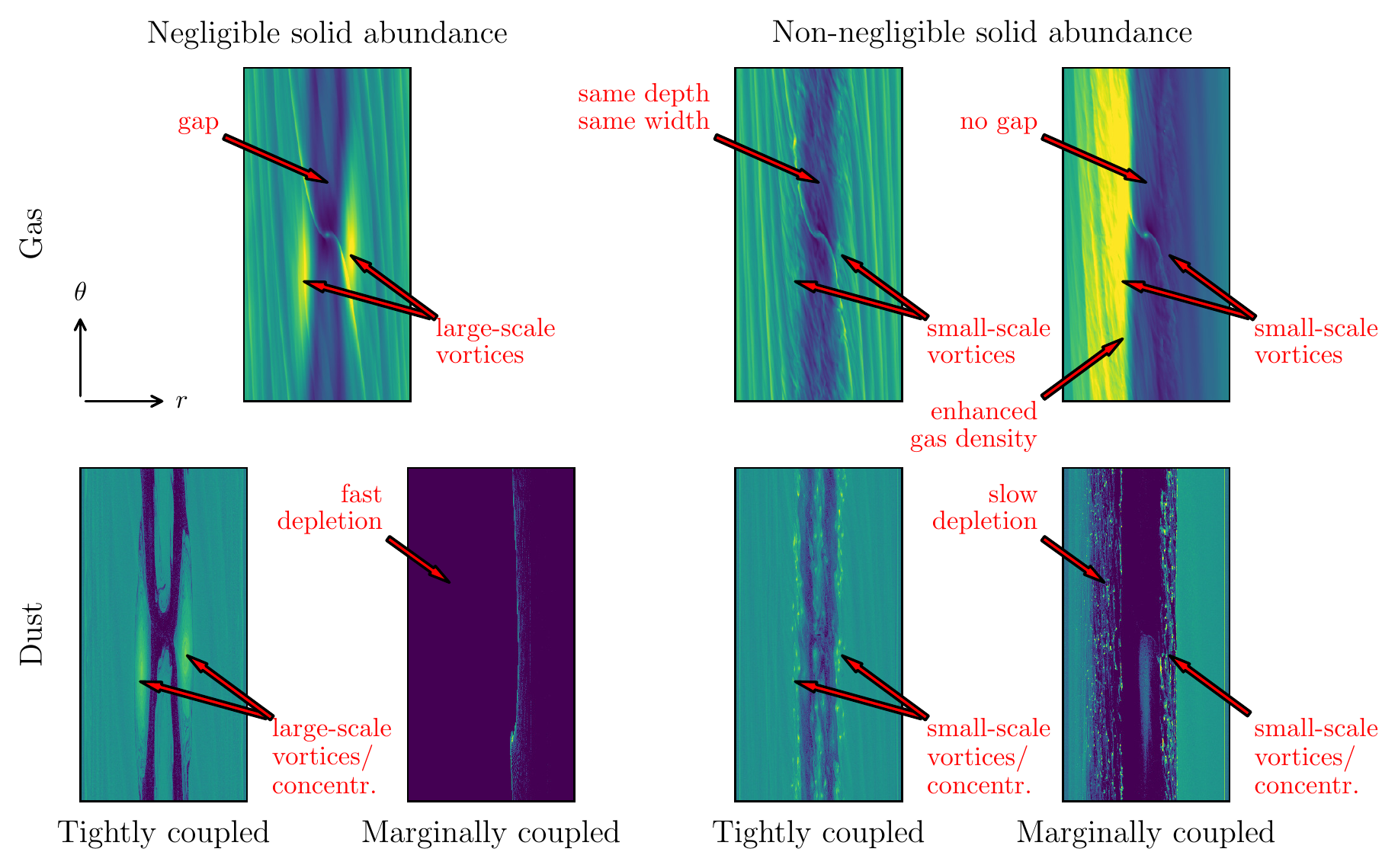}}\\
\caption{Schematic diagrams summarising the morphological signatures induced by dust back reaction in a dusty disc with an embedded planet.
We roughly define the dust particles as tightly coupled to the gas when the dimensionless stopping time $\taus < 0.1$ while as marginally coupled when $0.1 \lesssim \taus \lesssim 1$.}
\label{F:summary}
\end{center}
\end{figure*}

In Fig.~\ref{F:summary}, we schematically compile the morphological signatures induced by dust back reaction in a dusty disc with an embedded planet.
Figure~\ref{F:summary1} shows the results for a planet mass ($\Mp$) appreciably less than a thermal mass ($\Mth$; see equation~\eqref{E:mth}), while Fig.~\ref{F:summary2} shows those for a planet mass on the order of or more than a thermal mass.
We discuss in this section their implications for observations, while leaving the full summary of the effects by dust back reaction in Section~\ref{S:summary}.

First, the spiral structure is usually observed in the polarised light scattered from the $\micron$-sized dust particles near the surface of a disc.
It is believed that these particles are tightly coupled to the gas and trace the structure of the gas.
By comparing Figs.~\ref{F:sigma1}--\ref{F:sigma3}, one could see that the more massive the planet, the more open the spiral arms are (see Section~\ref{S:nbr}).
On the other hand, as discussed in Section~\ref{SSS:spiral1}, the more abundant the tightly coupled dust particles are, the less open the spiral arms become.
Therefore, there exists a degeneracy between the planet mass $\Mp$ and the solid abundance $Z$ of small dust particles when interpreting the spiral structure.
For example, the pitch angle increases by about 4$\degr$ at a distance of $x = 3\Hg$ from the planet when $\Mp$ increases from 0.1$\Mth$ to 1$\Mth$ \citep{BZ18}, while Fig.~\ref{F:spiral1} (or equation~\eqref{E:spiral}) shows that the pitch angle decreases by about 3$\fdg$6 when $Z$ increases from zero to unity.
This demonstrates that care needs to be taken in order to break this degeneracy by using other diagnostics.

Second, the interpretation of the gap structure in a disc is significantly more complicated.
A gap in the dust component becomes shallower with increasing solid abundance irrespective of the planet mass, while a gap in the gas component becomes deeper with increasing solid abundance only when the planet mass is small.
The width of a gas gap does not appear to be sensitive to the solid abundance.
On the other hand, it may be difficult to interpret the width of a dust gap separated by two axisymmetric rings, especially when the dust particles are marginally coupled.
If the solid abundance is high \emph{and} the planet is small (bottom-right panels of Fig.~\ref{F:summary1}), one might infer the existence of a planet near the centre of the gap.
However, if the solid abundance is low \emph{or} the planet is massive (bottom-left panels of Fig.~\ref{F:summary1} and bottom panels of Fig.~\ref{F:summary2}), the dust particles could drift radially inward inside the planetary orbit and accumulate at the outside of another planet (one should also consider that the higher the solid abundance, the slower the radial-drift speed and hence the dust depletion inside a planetary orbit; bottom-right panel of Fig.~\ref{F:summary2}).
In this case, two rings indicate the existence of \emph{two} planets.
Finally, a gap found in the dust component may not be directly translated to a gap in the gas component, except for the case of tightly coupled dust particles, as shown in the top-right panels of Figs.~\ref{F:summary1} and~\ref{F:summary2}.
When the solid abundance is high, a well-defined gap in the gas component may not exist.

Third, in the inviscid limit, the dust particles tends to form axisymmetric structure of narrow rings when they accumulate at a pressure maximum or when being pushed aside by a planet.
Even though a dust ring could be broken into small-scale vortical concentrations when the dust back reaction is in effect, as shown in Fig.~\ref{F:summary1}, it may still appear axisymmetric in observations.
Therefore, some kinematical information along a dust ring may be useful in detecting these small-scale vortices.
By contrast, a dust ring could become broad when the dust diffusion is appreciable \citep{KM18}, which could serve as an additional constraint on the strength of turbulence in observed discs \citep[e.g.,][]{DB18}.

Last but not least, a lopsided large-scale vortex may be unstable to dust back reaction, as shown in Sections~\ref{SSS:vortex1} and~\ref{SSS:vortex2}.
On the other hand, we demonstrate in Section~\ref{SSS:vortex2} a large-scale vortex may reappear after the dust particles are depleted.
Therefore, when a lopsided large-scale vortex is detected in the gas, the dust content might be low whether or not a lopsided concentration of dust particles is also detected.
This result may have important implications when interpreting the large-scale lopsided structure observed in some nearby protoplanetary discs like those around HD~142527 \citep{CP13} and Oph~IRS~48 \citep{MD13}.

We caution that the solid abundance $Z$ defined in this work (Section~\ref{S:method}) should not be taken literally as the ratio of the column densities when interpreting observations.
The dust particles sediment vertically and the dust disc can be one or even two orders of magnitude thinner than their gas counterpart \citep*{DMS95,YL07}, and the solid-to-gas density ratio $\epsilon$ near the mid-plane increases proportionately.
The gas drag is a local frictional force and hence the strength of the dust back reaction depends locally on $\epsilon$ instead of $Z$.
Because our models assume a razor-thin disc, therefore, we do not capture this 3D effect (see \citealt{KU17} and \citealt{GB19} for the case of viscous discs, however).
In this regard, the resulting dynamics in our models may be closer to that of a correspondingly lower $Z$.
We further discuss this caveat of our models in Section~\ref{S:summary}.

\section{Summary and Concluding Remarks} \label{S:summary}

In this work, we use the \pc{} and conduct a series of local-shearing-sheet simulations for an isothermal, inviscid, non-self-gravitating, razor-thin protoplanetary dust-gas disc under the influence of an embedded planet on a fixed circular orbit.
We systematically scan the parameter space by varying the planet mass ($0.1\Mth \le \Mp \le 1\Mth$), the stopping time of the mutual drag force between each dust particle and its surrounding gas ($10^{-3} \le \taus \le 1$), and the solid abundance ($0 < Z \le 1$).
Most of the models have the back reaction of the dust particles to the gas drag in effect, while some models turn off the back reaction (i.e., considering the limit of $Z \rightarrow 0$) for comparison purposes.
We pay specific attention to the morphological signatures in both the gas and the dust components of the disc induced by the dust back reaction, especially to the spiral structure, the gap structure, and the vortices.

We summarise in the following as well as in Fig.~\ref{F:summary} the resulting major features we have observed in our models.
\begin{itemize}
\item
The spiral structure may differ with different solid abundance $Z$, depending on how tightly the dust particles are coupled to the gas (Fig.~\ref{F:spiral}; Sections~\ref{SSS:spiral1} and~\ref{SSS:spiral2}).
When the disc consists of tightly coupled dust particles, the higher $Z$, the less open the spiral arms induced by the planet is.
This behaviour can be explained by the effective speed of sound modified by the dust back reaction in the approximation of a single dust-gas mixture (equations~\eqref{E:spiral}--\eqref{E:ess}).
On the contrary, the spiral structure is insensitive to $Z$ when the disc consists of marginally coupled dust particles, and the approximation of the single dust-gas mixture appears breaking down.
\item
A gap structure in the gas component of the disc driven by a planet can become extinguished due to the dust back reaction, especially when the disc has a significant amount of marginally coupled dust particles (Section~\ref{SSS:gap2}).
Because of the conservation of angular momentum, the gas radially drifts outward when the dust particles drift inward, and the gap structure becomes increasingly skewed with increasing angular momentum exchange between dust and gas.
The critical point occurs where the local solid-to-gas density ratio $\epsilon \sim 1$ and the dimensionless stopping time $\taus \sim 1$.
In this case, a local pressure maximum radially outside of the planet may not exist and hence the dust particles, irrespective of their sizes, can continue to drift past the planetary orbit unless the planet mass is on the order of or more than a thermal mass.
In other words, dust filtration by a low-mass ($\Mp < \Mth$) planet \citep{ZN12} could be reduced or even eliminated by dust back reaction.
We note also that this process -- which allows pebble accretion onto a proto-planet continue to operate -- may be another important factor in determining the pebble isolation mass (\citealt{MN12}; \citealt*{LJM14}; \citealt{BM18}).
\item
While the gap structure remains well defined in the case of tightly coupled dust particles, its dimensions change with the solid abundance $Z$ (Section~\ref{SSS:gap1}).
Unless the planet mass $\Mp \gtrsim \Mth$, the higher $Z$, the faster the gap in the gas component driven by the planet becomes deeper, which is also consistent with a reduced speed of sound in the approximation of a single dust-gas mixture (equation~\eqref{E:ess}).
However, the gap in the dust component shows the opposite trend when $\Mp \gtrsim 0.3\Mth$ because of the dust diffusion induced by dust-gas vortices near the gap edge.
On the other hand, the width of the gap structure does not appear to be sensitive to $Z$.
\item
The edges on both sides of the planet where the dust particles accumulate can become unstable and result in numerous \emph{small-scale} dust-gas vortices (Sections~\ref{SSS:vortex1} and~\ref{SSS:vortex2}).
The higher the solid abundance $Z$, the stronger this effect becomes.
We note that the dust accumulation near the outer edge is due to the radial drift toward a local pressure maximum while the one near the inner edge is due to the planetary torque.
Moreover, it appears that when the dust particles are tightly coupled to the gas, the system tends to regulate itself such that $\Sigmap / Z\Sigmag \simeq 2$ at the gap edge.
When the particles are marginally coupled, on the other hand, $\Sigmap \simeq 0.2\Sigma_{\text{g},0}$ at its peak.
\item
The large-scale vortices generated near the gap edges driven by a massive planet are broken into numerous \emph{small-scale} dust-gas vortices when the dust back reaction becomes significant (Fig.~\ref{F:sigma3}; Sections~\ref{SSS:vortex1} and~\ref{SSS:vortex2}).
\end{itemize}

As final remarks, several caveats need to be discussed here.
First, our models assume the inviscid limit.
The major effect of viscosity is to spread out the gaseous component of a disc and drive gas accretion onto the host star \citep{HC98}, the process of which cannot be captured by our approach of the local-shearing-sheet approximation.
Current MHD models, however, indicate that accretion predominantly occurs near the surface of a disc \citep[e.g.,][]{BS13,ZS18}, which is significantly away from the mid-plane where most mm/cm-sized dust particles are located.
Moreover, current observations show that the magnitude of the turbulent velocity at any given height in a protoplanetary disc remains below a few percent of the speed of sound, implying a low turbulent viscosity \citep{FH17,FH18}.
In any case, viscosity in the context of planet-disc interaction tends to balance the planetary torque and hence an equilibrium gap structure in the gas component of the disc may be established \citep[see, e.g.,][]{KT15}.

On the other hand, there exists some general confusion between viscosity discussed above and turbulent diffusion of dust particles.
Turbulence near the mid-plane of a disc driven by non-ideal MHD can be highly anisotropic and the magnitude of viscous stress can significantly differ from that of turbulent diffusion of dust particles (\citealt*{ZSB15}; \citealt{RL18}; \citealt{YMJ18}).
The dimensionless diffusion coefficients for mm/cm-sized dust particles inferred by recent ALMA observations are on the order of $\alpha_\text{t} \sim 10^{-4}$ \citep{PD16,DB18}.
It is unclear how turbulent diffusion of this magnitude influences the results found in this work.

Second, the length scale of a morphological feature in a disc -- such as spiral arms, rings, gaps, and vortices -- is in general proportional to the speed of sound $\cs$ \citep{GR01} and hence depends on the equation of state.
Because $\cs \propto \sqrt{\gamma}$, where $\gamma$ is the adiabatic index, the spiral arms tend to be more open and the gaps wider and shallower with increasing $\gamma$ in comparison with models assuming an isothermal equation of state as studied in this work \citep{DZ15}.
Moreover, it appears that the spiral and the gap structures are sensitive to the cooling time-scale of the disc, especially when it is comparable with the orbital period \citep{MR19a,MR19b,ZZ19}.
It remains to be seen whether or not the effects of dust back reaction presented in this work can be considered independent of the aforementioned properties.

Third, our models assume a razor-thin disc and ignore its vertical extension.
With a vertically resolved scale height of the disc, however, the disc is subject to the streaming instability \citep{YG05}.
The non-linear evolution of the instability in full 3D \citep[e.g.,][]{YJ14} certainly interacts with all the dust-gas dynamics we have observed in this work.

Finally, our models assume the dust component of the disc consists of identical particles.
More realistically, however, a disc should consist of dust particles of a wide range of sizes and properties, and they indirectly interact with each other via the dust back reaction to the gas (\citealt{NSH86}; \citealt{BS10}; \citealt{GLM17}; \citealt*{SYJ18}; \citealt{KB19}), let alone including the process of dust coagulation and fragmentation \citep*[see, e.g.,][]{BFJ16}.
We note that recently \cite{DL19} simulated a planet-induced gap in a turbulent disc with a full size distribution of dust and its coagulation, and concluded that the dust back reaction does not play an important role.
A potential reason is that the adopted solid abundance $Z = 0.01$ was low and the effect of vertical sedimentation was not considered (see the discussion in the end of Section~\ref{S:obs}).
On the other hand, \cite{KU17} and \cite{GB19} did use a vertically integrated approach and found that the dust back reaction is not important unless the solid abundance is relatively high ($Z \gtrsim 0.03$).
Therefore, except for the effects of turbulent diffusion of dust particles discussed above, our results using relatively high $Z$s may remain consistent.

We believe the above considerations along with dust back reaction should be some of the major topics in understanding the observations of the protoplanetary discs in future investigations.

\section*{Acknowledgements}

We thank the anonymous reviewer for the useful comments.
We also appreciate the discussion of this work with Giovanni Dipierro, Ruobing Dong, Joanna Dr{\k{a}}{\.z}kowska, Min-Kai Lin, and Wladimir Lyra.
This work used the Extreme Science and Engineering Discovery Environment (XSEDE) Stampede2 at the Texas Advanced Computing Center (TACC) through allocation AST130002.
We acknowledge support from the National Science Foundation under CAREER grant No.~AST-1753168, Sloan Research Fellowship, and the National Aeronautics and Space Administration through the Astrophysics Theory Program with grant No.~NNX17AK40G.

\bibliographystyle{mnras}
\bibliography{ms}

\begin{thebibliography}{}
\makeatletter
\relax
\def\mn@urlcharsother{\let\do\@makeother \do\$\do\&\do\#\do\^\do\_\do\%\do\~}
\def\mn@doi{\begingroup\mn@urlcharsother \@ifnextchar [ {\mn@doi@}
  {\mn@doi@[]}}
\def\mn@doi@[#1]#2{\def\@tempa{#1}\ifx\@tempa\@empty \href
  {http://dx.doi.org/#2} {doi:#2}\else \href {http://dx.doi.org/#2} {#1}\fi
  \endgroup}
\def\mn@eprint#1#2{\mn@eprint@#1:#2::\@nil}
\def\mn@eprint@arXiv#1{\href {http://arxiv.org/abs/#1} {{\tt arXiv:#1}}}
\def\mn@eprint@dblp#1{\href {http://dblp.uni-trier.de/rec/bibtex/#1.xml}
  {dblp:#1}}
\def\mn@eprint@#1:#2:#3:#4\@nil{\def\@tempa {#1}\def\@tempb {#2}\def\@tempc
  {#3}\ifx \@tempc \@empty \let \@tempc \@tempb \let \@tempb \@tempa \fi \ifx
  \@tempb \@empty \def\@tempb {arXiv}\fi \@ifundefined
  {mn@eprint@\@tempb}{\@tempb:\@tempc}{\expandafter \expandafter \csname
  mn@eprint@\@tempb\endcsname \expandafter{\@tempc}}}

\bibitem[\protect\citeauthoryear{{ALMA Partnership} et~al.,}{{ALMA Partnership}
  et~al.}{2015}]{AB15}
{ALMA Partnership} et~al., 2015, \mn@doi [\apjl] {10.1088/2041-8205/808/1/L3},
  \href {https://ui.adsabs.harvard.edu/abs/2015ApJ...808L...3A} {808, L3}

\bibitem[\protect\citeauthoryear{{Andrews} et~al.,}{{Andrews}
  et~al.}{2018}]{AH18}
{Andrews} S.~M.,  et~al., 2018, \mn@doi [\apjl] {10.3847/2041-8213/aaf741},
  \href {https://ui.adsabs.harvard.edu/abs/2018ApJ...869L..41A} {869, L41}

\bibitem[\protect\citeauthoryear{{Bae} \& {Zhu}}{{Bae} \& {Zhu}}{2018}]{BZ18}
{Bae} J.,  {Zhu} Z.,  2018, \mn@doi [\apj] {10.3847/1538-4357/aabf8c}, \href
  {https://ui.adsabs.harvard.edu/abs/2018ApJ...859..118B} {859, 118}

\bibitem[\protect\citeauthoryear{{Bae}, {Zhu}  \& {Hartmann}}{{Bae}
  et~al.}{2016}]{BZH16}
{Bae} J.,  {Zhu} Z.,   {Hartmann} L.,  2016, \mn@doi [\apj]
  {10.3847/0004-637X/819/2/134}, \href
  {https://ui.adsabs.harvard.edu/abs/2016ApJ...819..134B} {819, 134}

\bibitem[\protect\citeauthoryear{{Bae}, {Zhu}  \& {Hartmann}}{{Bae}
  et~al.}{2017}]{BZH17}
{Bae} J.,  {Zhu} Z.,   {Hartmann} L.,  2017, \mn@doi [\apj]
  {10.3847/1538-4357/aa9705}, \href
  {https://ui.adsabs.harvard.edu/abs/2017ApJ...850..201B} {850, 201}

\bibitem[\protect\citeauthoryear{{Bai} \& {Stone}}{{Bai} \&
  {Stone}}{2010}]{BS10}
{Bai} X.-N.,  {Stone} J.~M.,  2010, \mn@doi [\apj]
  {10.1088/0004-637X/722/2/1437}, \href
  {https://ui.adsabs.harvard.edu/abs/2010ApJ...722.1437B} {722, 1437}

\bibitem[\protect\citeauthoryear{{Bai} \& {Stone}}{{Bai} \&
  {Stone}}{2013}]{BS13}
{Bai} X.-N.,  {Stone} J.~M.,  2013, \mn@doi [\apj]
  {10.1088/0004-637X/769/1/76}, \href
  {https://ui.adsabs.harvard.edu/abs/2013ApJ...769...76B} {769, 76}

\bibitem[\protect\citeauthoryear{{Ben{\'\i}tez-Llambay} \&
  {Pessah}}{{Ben{\'\i}tez-Llambay} \& {Pessah}}{2018}]{BP18}
{Ben{\'\i}tez-Llambay} P.,  {Pessah} M.~E.,  2018, \mn@doi [\apj]
  {10.3847/2041-8213/aab2ae}, \href
  {https://ui.adsabs.harvard.edu/abs/2018ApJ...855L..28B} {855, L28}

\bibitem[\protect\citeauthoryear{{Birnstiel}, {Fang}  \&
  {Johansen}}{{Birnstiel} et~al.}{2016}]{BFJ16}
{Birnstiel} T.,  {Fang} M.,   {Johansen} A.,  2016, \mn@doi [\ssr]
  {10.1007/s11214-016-0256-1}, \href
  {https://ui.adsabs.harvard.edu/abs/2016SSRv..205...41B} {205, 41}

\bibitem[\protect\citeauthoryear{{Bitsch}, {Johansen}, {Lambrechts}  \&
  {Morbidelli}}{{Bitsch} et~al.}{2015}]{BJ15}
{Bitsch} B.,  {Johansen} A.,  {Lambrechts} M.,   {Morbidelli} A.,  2015,
  \mn@doi [\aap] {10.1051/0004-6361/201424964}, \href
  {https://ui.adsabs.harvard.edu/\#abs/2015A&A...575A..28B} {575, A28}

\bibitem[\protect\citeauthoryear{{Bitsch}, {Morbidelli}, {Johansen}, {Lega},
  {Lambrechts}  \& {Crida}}{{Bitsch} et~al.}{2018}]{BM18}
{Bitsch} B.,  {Morbidelli} A.,  {Johansen} A.,  {Lega} E.,  {Lambrechts} M.,
  {Crida} A.,  2018, \mn@doi [\aap] {10.1051/0004-6361/201731931}, \href
  {https://ui.adsabs.harvard.edu/abs/2018A&A...612A..30B} {612, A30}

\bibitem[\protect\citeauthoryear{{Brandenburg} \& {Dobler}}{{Brandenburg} \&
  {Dobler}}{2002}]{BD02}
{Brandenburg} A.,  {Dobler} W.,  2002, \mn@doi [Computer Physics
  Communications] {10.1016/S0010-4655(02)00334-X}, \href
  {https://ui.adsabs.harvard.edu/\#abs/2002CoPhC.147..471B} {147, 471}

\bibitem[\protect\citeauthoryear{{Brandenburg}, {Nordlund}, {Stein}  \&
  {Torkelsson}}{{Brandenburg} et~al.}{1995}]{BN95}
{Brandenburg} A.,  {Nordlund} A.,  {Stein} R.~F.,   {Torkelsson} U.,  1995,
  \mn@doi [\apj] {10.1086/175831}, \href
  {https://ui.adsabs.harvard.edu/\#abs/1995ApJ...446..741B} {446, 741}

\bibitem[\protect\citeauthoryear{{Casassus} et~al.,}{{Casassus}
  et~al.}{2013}]{CP13}
{Casassus} S.,  et~al., 2013, \mn@doi [\nat] {10.1038/nature11769}, \href
  {https://ui.adsabs.harvard.edu/abs/2013Natur.493..191C} {493, 191}

\bibitem[\protect\citeauthoryear{{Castrejon}, {Lyra}, {Richert}  \&
  {Kuchner}}{{Castrejon} et~al.}{2019}]{CL19}
{Castrejon} A.,  {Lyra} W.,  {Richert} A. J.~W.,   {Kuchner} M.,  2019,
  \apj, submitted
  (\href {https://arxiv.org/abs/1906.04816} {arXiv:1906.04816})

\bibitem[\protect\citeauthoryear{{Chen} \& {Lin}}{{Chen} \& {Lin}}{2018}]{CL18}
{Chen} J.-W.,  {Lin} M.-K.,  2018, \mn@doi [\mnras] {10.1093/mnras/sty1166},
  \href {https://ui.adsabs.harvard.edu/abs/2018MNRAS.478.2737C} {478, 2737}

\bibitem[\protect\citeauthoryear{{Crnkovic-Rubsamen}, {Zhu}  \&
  {Stone}}{{Crnkovic-Rubsamen} et~al.}{2015}]{CZS15}
{Crnkovic-Rubsamen} I.,  {Zhu} Z.,   {Stone} J.~M.,  2015, \mn@doi [\mnras]
  {10.1093/mnras/stv828}, \href
  {https://ui.adsabs.harvard.edu/abs/2015MNRAS.450.4285C} {450, 4285}

\bibitem[\protect\citeauthoryear{{Dipierro}, {Price}, {Laibe}, {Hirsh},
  {Cerioli}  \& {Lodato}}{{Dipierro} et~al.}{2015}]{DP15}
{Dipierro} G.,  {Price} D.,  {Laibe} G.,  {Hirsh} K.,  {Cerioli} A.,   {Lodato}
  G.,  2015, \mn@doi [\mnras] {10.1093/mnrasl/slv105}, \href
  {https://ui.adsabs.harvard.edu/abs/2015MNRAS.453L..73D} {453, L73}

\bibitem[\protect\citeauthoryear{{Dong}, {Rafikov}, {Stone}  \&
  {Petrovich}}{{Dong} et~al.}{2011a}]{DR11}
{Dong} R.,  {Rafikov} R.~R.,  {Stone} J.~M.,   {Petrovich} C.,  2011a, \mn@doi
  [\apj] {10.1088/0004-637X/741/1/56}, \href
  {https://ui.adsabs.harvard.edu/\#abs/2011ApJ...741...56D} {741, 56}

\bibitem[\protect\citeauthoryear{{Dong}, {Rafikov}  \& {Stone}}{{Dong}
  et~al.}{2011b}]{DRS11}
{Dong} R.,  {Rafikov} R.~R.,   {Stone} J.~M.,  2011b, \mn@doi [\apj]
  {10.1088/0004-637X/741/1/57}, \href
  {https://ui.adsabs.harvard.edu/\#abs/2011ApJ...741...57D} {741, 57}

\bibitem[\protect\citeauthoryear{{Dong}, {Zhu}  \& {Whitney}}{{Dong}
  et~al.}{2015a}]{DZW15}
{Dong} R.,  {Zhu} Z.,   {Whitney} B.,  2015a, \mn@doi [\apj]
  {10.1088/0004-637X/809/1/93}, \href
  {https://ui.adsabs.harvard.edu/abs/2015ApJ...809...93D} {809, 93}

\bibitem[\protect\citeauthoryear{{Dong}, {Zhu}, {Rafikov}  \& {Stone}}{{Dong}
  et~al.}{2015b}]{DZ15}
{Dong} R.,  {Zhu} Z.,  {Rafikov} R.~R.,   {Stone} J.~M.,  2015b, \mn@doi
  [\apjl] {10.1088/2041-8205/809/1/L5}, \href
  {https://ui.adsabs.harvard.edu/abs/2015ApJ...809L...5D} {809, L5}

\bibitem[\protect\citeauthoryear{{Dong}, {Li}, {Chiang}  \& {Li}}{{Dong}
  et~al.}{2017}]{DLC17}
{Dong} R.,  {Li} S.,  {Chiang} E.,   {Li} H.,  2017, \mn@doi [\apj]
  {10.3847/1538-4357/aa72f2}, \href
  {https://ui.adsabs.harvard.edu/abs/2017ApJ...843..127D} {843, 127}

\bibitem[\protect\citeauthoryear{{Dong} et~al.,}{{Dong} et~al.}{2018a}]{DL18}
{Dong} R.,  et~al., 2018a, \mn@doi [\apj] {10.3847/1538-4357/aac6cb}, \href
  {https://ui.adsabs.harvard.edu/abs/2018ApJ...860..124D} {860, 124}

\bibitem[\protect\citeauthoryear{{Dong}, {Li}, {Chiang}  \& {Li}}{{Dong}
  et~al.}{2018b}]{DLC18}
{Dong} R.,  {Li} S.,  {Chiang} E.,   {Li} H.,  2018b, \mn@doi [\apj]
  {10.3847/1538-4357/aadadd}, \href
  {https://ui.adsabs.harvard.edu/abs/2018ApJ...866..110D} {866, 110}

\bibitem[\protect\citeauthoryear{{Dr{\k{a}}{\.z}kowska}, {Li}, {Birnstiel},
  {Stammler}  \& {Li}}{{Dr{\k{a}}{\.z}kowska} et~al.}{2019}]{DL19}
{Dr{\k{a}}{\.z}kowska} J.,  {Li} S.,  {Birnstiel} T.,  {Stammler} S.~M.,   {Li}
  H.,  2019, \apj, in press
  (\href {https://arxiv.org/abs/1909.10526} {arXiv:1909.10526})

\bibitem[\protect\citeauthoryear{{Dubrulle}, {Morfill}  \&
  {Sterzik}}{{Dubrulle} et~al.}{1995}]{DMS95}
{Dubrulle} B.,  {Morfill} G.,   {Sterzik} M.,  1995, \mn@doi [\icarus]
  {10.1006/icar.1995.1058}, \href
  {https://ui.adsabs.harvard.edu/abs/1995Icar..114..237D} {114, 237}

\bibitem[\protect\citeauthoryear{{Dullemond} et~al.,}{{Dullemond}
  et~al.}{2018}]{DB18}
{Dullemond} C.~P.,  et~al., 2018, \mn@doi [\apjl] {10.3847/2041-8213/aaf742},
  \href {https://ui.adsabs.harvard.edu/abs/2018ApJ...869L..46D} {869, L46}

\bibitem[\protect\citeauthoryear{{Flaherty} et~al.,}{{Flaherty}
  et~al.}{2017}]{FH17}
{Flaherty} K.~M.,  et~al., 2017, \mn@doi [\apj] {10.3847/1538-4357/aa79f9},
  \href {https://ui.adsabs.harvard.edu/abs/2017ApJ...843..150F} {843, 150}

\bibitem[\protect\citeauthoryear{{Flaherty}, {Hughes}, {Teague}, {Simon},
  {Andrews}  \& {Wilner}}{{Flaherty} et~al.}{2018}]{FH18}
{Flaherty} K.~M.,  {Hughes} A.~M.,  {Teague} R.,  {Simon} J.~B.,  {Andrews}
  S.~M.,   {Wilner} D.~J.,  2018, \mn@doi [\apj] {10.3847/1538-4357/aab615},
  \href {https://ui.adsabs.harvard.edu/abs/2018ApJ...856..117F} {856, 117}

\bibitem[\protect\citeauthoryear{{Fu}, {Li}, {Lubow}, {Li}  \& {Liang}}{{Fu}
  et~al.}{2014}]{FL14}
{Fu} W.,  {Li} H.,  {Lubow} S.,  {Li} S.,   {Liang} E.,  2014, \mn@doi [\apjl]
  {10.1088/2041-8205/795/2/L39}, \href
  {https://ui.adsabs.harvard.edu/abs/2014ApJ...795L..39F} {795, L39}

\bibitem[\protect\citeauthoryear{{G{\'a}rate}, {Birnstiel},
  {Dr{\k{a}}{\.z}kowska}  \& {Stammler}}{{G{\'a}rate} et~al.}{2019}]{GB19}
{G{\'a}rate} M.,  {Birnstiel} T.,  {Dr{\k{a}}{\.z}kowska} J.,   {Stammler}
  S.~M.,  2019, \aap, submitted
  (\href {https://arxiv.org/abs/1906.07708} {arXiv:1906.07708})

\bibitem[\protect\citeauthoryear{{Garufi} et~al.,}{{Garufi}
  et~al.}{2017}]{GB17}
{Garufi} A.,  et~al., 2017, \mn@doi [The Messenger] {10.18727/0722-6691/5036},
  \href {https://ui.adsabs.harvard.edu/abs/2017Msngr.169...32G} {169, 32}

\bibitem[\protect\citeauthoryear{{Garufi} et~al.,}{{Garufi}
  et~al.}{2018}]{GB18}
{Garufi} A.,  et~al., 2018, \mn@doi [\aap] {10.1051/0004-6361/201833872}, \href
  {https://ui.adsabs.harvard.edu/abs/2018A&A...620A..94G} {620, A94}

\bibitem[\protect\citeauthoryear{{Gent}, {Mac Low}, {K{\"a}pyl{\"a}}, {Sarson}
  \& {Hollins}}{{Gent} et~al.}{2019}]{GM18}
{Gent} F.~A.,  {Mac Low} M.-M.,  {K{\"a}pyl{\"a}} M.~J.,  {Sarson} G.~R.,
  {Hollins} J.~F., 2019, Geophysical \& Astrophysical Fluid Dynamics,
  doi:~\href {https://doi.org/10.1080/03091929.2019.1634705}
                             {10.1080/03091929.2019.1634705}

\bibitem[\protect\citeauthoryear{{Goldreich} \& {Lynden-Bell}}{{Goldreich} \&
  {Lynden-Bell}}{1965}]{GL65}
{Goldreich} P.,  {Lynden-Bell} D.,  1965, \mn@doi [\mnras]
  {10.1093/mnras/130.2.125}, \href
  {https://ui.adsabs.harvard.edu/\#abs/1965MNRAS.130..125G} {130, 125}

\bibitem[\protect\citeauthoryear{{Goldreich} \& {Tremaine}}{{Goldreich} \&
  {Tremaine}}{1979}]{GT79}
{Goldreich} P.,  {Tremaine} S.,  1979, \mn@doi [\apj] {10.1086/157448}, \href
  {https://ui.adsabs.harvard.edu/\#abs/1979ApJ...233..857G} {233, 857}

\bibitem[\protect\citeauthoryear{{Goldreich} \& {Tremaine}}{{Goldreich} \&
  {Tremaine}}{1980}]{GT80}
{Goldreich} P.,  {Tremaine} S.,  1980, \mn@doi [\apj] {10.1086/158356}, \href
  {https://ui.adsabs.harvard.edu/\#abs/1980ApJ...241..425G} {241, 425}

\bibitem[\protect\citeauthoryear{{Gonzalez}, {Laibe}  \& {Maddison}}{{Gonzalez}
  et~al.}{2017}]{GLM17}
{Gonzalez} J.~F.,  {Laibe} G.,   {Maddison} S.~T.,  2017, \mn@doi [\mnras]
  {10.1093/mnras/stx016}, \href
  {https://ui.adsabs.harvard.edu/abs/2017MNRAS.467.1984G} {467, 1984}

\bibitem[\protect\citeauthoryear{{Goodman} \& {Rafikov}}{{Goodman} \&
  {Rafikov}}{2001}]{GR01}
{Goodman} J.,  {Rafikov} R.~R.,  2001, \mn@doi [\apj] {10.1086/320572}, \href
  {https://ui.adsabs.harvard.edu/\#abs/2001ApJ...552..793G} {552, 793}

\bibitem[\protect\citeauthoryear{{Hartmann}, {Calvet}, {Gullbring}  \&
  {D'Alessio}}{{Hartmann} et~al.}{1998}]{HC98}
{Hartmann} L.,  {Calvet} N.,  {Gullbring} E.,   {D'Alessio} P.,  1998, \mn@doi
  [\apj] {10.1086/305277}, \href
  {https://ui.adsabs.harvard.edu/abs/1998ApJ...495..385H} {495, 385}

\bibitem[\protect\citeauthoryear{{Hawley}, {Gammie}  \& {Balbus}}{{Hawley}
  et~al.}{1995}]{HGB95}
{Hawley} J.~F.,  {Gammie} C.~F.,   {Balbus} S.~A.,  1995, \mn@doi [\apj]
  {10.1086/175311}, \href
  {https://ui.adsabs.harvard.edu/\#abs/1995ApJ...440..742H} {440, 742}

\bibitem[\protect\citeauthoryear{{Hockney} \& {Eastwood}}{{Hockney} \&
  {Eastwood}}{1988}]{HE88}
{Hockney} R.~W.,  {Eastwood} J.~W.,  1988, {Computer Simulation Using
  Particles}.
{Taylor \& Francis, New York, NY}

\bibitem[\protect\citeauthoryear{{Johansen} \& {Lambrechts}}{{Johansen} \&
  {Lambrechts}}{2017}]{JL17}
{Johansen} A.,  {Lambrechts} M.,  2017, \mn@doi [Annual Review of Earth and
  Planetary Sciences] {10.1146/annurev-earth-063016-020226}, \href
  {https://ui.adsabs.harvard.edu/\#abs/2017AREPS..45..359J} {45, 359}

\bibitem[\protect\citeauthoryear{{Johansen}, {Blum}, {Tanaka}, {Ormel},
  {Bizzarro}  \& {Rickman}}{{Johansen} et~al.}{2014}]{JB14}
{Johansen} A.,  {Blum} J.,  {Tanaka} H.,  {Ormel} C.,  {Bizzarro} M.,
  {Rickman} H.,  2014, in {Beuther} H.,  {Klessen} R.~S.,  {Dullemond} C.~P.,
  {Henning} T.,  eds, Protostars and Planets VI. p.~547 (\mn@eprint {arXiv}
  {1402.1344}), \mn@doi{10.2458/azu_uapress_9780816531240-ch024}

\bibitem[\protect\citeauthoryear{{Kanagawa}, {Tanaka}, {Muto}, {Tanigawa}  \&
  {Takeuchi}}{{Kanagawa} et~al.}{2015}]{KT15}
{Kanagawa} K.~D.,  {Tanaka} H.,  {Muto} T.,  {Tanigawa} T.,   {Takeuchi} T.,
  2015, \mn@doi [\mnras] {10.1093/mnras/stv025}, \href
  {https://ui.adsabs.harvard.edu/abs/2015MNRAS.448..994K} {448, 994}

\bibitem[\protect\citeauthoryear{{Kanagawa}, {Ueda}, {Muto}  \&
  {Okuzumi}}{{Kanagawa} et~al.}{2017}]{KU17}
{Kanagawa} K.~D.,  {Ueda} T.,  {Muto} T.,   {Okuzumi} S.,  2017, \mn@doi [\apj]
  {10.3847/1538-4357/aa7ca1}, \href
  {https://ui.adsabs.harvard.edu/abs/2017ApJ...844..142K} {844, 142}

\bibitem[\protect\citeauthoryear{{Kanagawa}, {Muto}, {Okuzumi}, {Tanigawa},
  {Taki}  \& {Shibaike}}{{Kanagawa} et~al.}{2018}]{KM18}
{Kanagawa} K.~D.,  {Muto} T.,  {Okuzumi} S.,  {Tanigawa} T.,  {Taki} T.,
  {Shibaike} Y.,  2018, \mn@doi [\apj] {10.3847/1538-4357/aae837}, \href
  {https://ui.adsabs.harvard.edu/abs/2018ApJ...868...48K} {868, 48}

\bibitem[\protect\citeauthoryear{{Krapp}, {Ben{\'\i}tez-Llambay}, {Gressel}  \&
  {Pessah}}{{Krapp} et~al.}{2019}]{KB19}
{Krapp} L.,  {Ben{\'\i}tez-Llambay} P.,  {Gressel} O.,   {Pessah} M.~E.,  2019,
  \mn@doi [\apjl] {10.3847/2041-8213/ab2596}, \href
  {https://ui.adsabs.harvard.edu/abs/2019ApJ...878L..30K} {878, L30}

\bibitem[\protect\citeauthoryear{{Laibe} \& {Price}}{{Laibe} \&
  {Price}}{2014}]{LP14}
{Laibe} G.,  {Price} D.~J.,  2014, \mn@doi [\mnras] {10.1093/mnras/stu355},
  \href {https://ui.adsabs.harvard.edu/abs/2014MNRAS.440.2136L} {440, 2136}

\bibitem[\protect\citeauthoryear{{Lambrechts}, {Johansen}  \&
  {Morbidelli}}{{Lambrechts} et~al.}{2014}]{LJM14}
{Lambrechts} M.,  {Johansen} A.,   {Morbidelli} A.,  2014, \mn@doi [\aap]
  {10.1051/0004-6361/201423814}, \href
  {https://ui.adsabs.harvard.edu/abs/2014A&A...572A..35L} {572, A35}

\bibitem[\protect\citeauthoryear{{Lin} \& {Youdin}}{{Lin} \&
  {Youdin}}{2017}]{LY17}
{Lin} M.-K.,  {Youdin} A.~N.,  2017, \mn@doi [\apj] {10.3847/1538-4357/aa92cd},
  \href {https://ui.adsabs.harvard.edu/abs/2017ApJ...849..129L} {849, 129}

\bibitem[\protect\citeauthoryear{{Long} et~al.,}{{Long} et~al.}{2018}]{LP18}
{Long} F.,  et~al., 2018, \mn@doi [\apj] {10.3847/1538-4357/aae8e1}, \href
  {https://ui.adsabs.harvard.edu/abs/2018ApJ...869...17L} {869, 17}

\bibitem[\protect\citeauthoryear{{Lovelace}, {Li}, {Colgate}  \&
  {Nelson}}{{Lovelace} et~al.}{1999}]{LL99}
{Lovelace} R.~V.~E.,  {Li} H.,  {Colgate} S.~A.,   {Nelson} A.~F.,  1999,
  \mn@doi [\apj] {10.1086/306900}, \href
  {https://ui.adsabs.harvard.edu/abs/1999ApJ...513..805L} {513, 805}

\bibitem[\protect\citeauthoryear{{Lyra} \& {Lin}}{{Lyra} \& {Lin}}{2013}]{LL13}
{Lyra} W.,  {Lin} M.-K.,  2013, \mn@doi [\apj] {10.1088/0004-637X/775/1/17},
  \href {https://ui.adsabs.harvard.edu/abs/2013ApJ...775...17L} {775, 17}

\bibitem[\protect\citeauthoryear{{Lyra}, {Johansen}, {Klahr}  \&
  {Piskunov}}{{Lyra} et~al.}{2009}]{LJ09}
{Lyra} W.,  {Johansen} A.,  {Klahr} H.,   {Piskunov} N.,  2009, \mn@doi [\aap]
  {10.1051/0004-6361:200810797}, \href
  {https://ui.adsabs.harvard.edu/abs/2009A&A...493.1125L} {493, 1125}

\bibitem[\protect\citeauthoryear{{Miranda} \& {Rafikov}}{{Miranda} \&
  {Rafikov}}{2019a}]{MR19a}
{Miranda} R.,  {Rafikov} R.~R.,  2019a, \mn@doi [\apjl]
  {10.3847/2041-8213/ab22a7}, \href
  {https://ui.adsabs.harvard.edu/abs/2019ApJ...878L...9M} {878, L9}

\bibitem[\protect\citeauthoryear{{Miranda} \& {Rafikov}}{{Miranda} \&
  {Rafikov}}{2019b}]{MR19b}
{Miranda} R.,  {Rafikov} R.~R.,  2019b, AAS Journals, submitted\\(\href
  {https://ui.adsabs.harvard.edu/abs/2019arXiv191101428M} {arXiv:1911.01428})

\bibitem[\protect\citeauthoryear{{Morbidelli} \& {Nesvorny}}{{Morbidelli} \&
  {Nesvorny}}{2012}]{MN12}
{Morbidelli} A.,  {Nesvorny} D.,  2012, \mn@doi [\aap]
  {10.1051/0004-6361/201219824}, \href
  {https://ui.adsabs.harvard.edu/abs/2012A&A...546A..18M} {546, A18}

\bibitem[\protect\citeauthoryear{{Nakagawa}, {Sekiya}  \& {Hayashi}}{{Nakagawa}
  et~al.}{1986}]{NSH86}
{Nakagawa} Y.,  {Sekiya} M.,   {Hayashi} C.,  1986, \mn@doi [\icarus]
  {10.1016/0019-1035(86)90121-1}, \href
  {https://ui.adsabs.harvard.edu/\#abs/1986Icar...67..375N} {67, 375}

\bibitem[\protect\citeauthoryear{{Ogilvie} \& {Lubow}}{{Ogilvie} \&
  {Lubow}}{2002}]{OL02}
{Ogilvie} G.~I.,  {Lubow} S.~H.,  2002, \mn@doi [\mnras]
  {10.1046/j.1365-8711.2002.05148.x}, \href
  {https://ui.adsabs.harvard.edu/abs/2002MNRAS.330..950O} {330, 950}

\bibitem[\protect\citeauthoryear{{Paardekooper}, {Baruteau}, {Crida}  \&
  {Kley}}{{Paardekooper} et~al.}{2010}]{PB10}
{Paardekooper} S.~J.,  {Baruteau} C.,  {Crida} A.,   {Kley} W.,  2010, \mn@doi
  [\mnras] {10.1111/j.1365-2966.2009.15782.x}, \href
  {https://ui.adsabs.harvard.edu/abs/2010MNRAS.401.1950P} {401, 1950}

\bibitem[\protect\citeauthoryear{{Pierens}, {Lin}  \& {Raymond}}{{Pierens}
  et~al.}{2019}]{PLR19}
{Pierens} A.,  {Lin} M.~K.,   {Raymond} S.~N.,  2019, \mn@doi [\mnras]
  {10.1093/mnras/stz1718}, \href
  {https://ui.adsabs.harvard.edu/abs/2019MNRAS.488..645P} {488, 645}

\bibitem[\protect\citeauthoryear{{Pinte}, {Dent}, {M{\'e}nard}, {Hales},
  {Hill}, {Cortes}  \& {de Gregorio-Monsalvo}}{{Pinte} et~al.}{2016}]{PD16}
{Pinte} C.,  {Dent} W.~R.~F.,  {M{\'e}nard} F.,  {Hales} A.,  {Hill} T.,
  {Cortes} P.,   {de Gregorio-Monsalvo} I.,  2016, \mn@doi [\apj]
  {10.3847/0004-637X/816/1/25}, \href
  {https://ui.adsabs.harvard.edu/abs/2016ApJ...816...25P} {816, 25}

\bibitem[\protect\citeauthoryear{{Raettig}, {Klahr}  \& {Lyra}}{{Raettig}
  et~al.}{2015}]{RKL15}
{Raettig} N.,  {Klahr} H.,   {Lyra} W.,  2015, \mn@doi [\apj]
  {10.1088/0004-637X/804/1/35}, \href
  {https://ui.adsabs.harvard.edu/abs/2015ApJ...804...35R} {804, 35}

\bibitem[\protect\citeauthoryear{{Riols} \& {Lesur}}{{Riols} \&
  {Lesur}}{2018}]{RL18}
{Riols} A.,  {Lesur} G.,  2018, \mn@doi [\aap] {10.1051/0004-6361/201833212},
  \href {https://ui.adsabs.harvard.edu/abs/2018A&A...617A.117R} {617, A117}

\bibitem[\protect\citeauthoryear{{Sch{\"a}fer}, {Yang}  \&
  {Johansen}}{{Sch{\"a}fer} et~al.}{2017}]{SYJ17}
{Sch{\"a}fer} U.,  {Yang} C.-C.,   {Johansen} A.,  2017, \mn@doi [\aap]
  {10.1051/0004-6361/201629561}, \href
  {https://ui.adsabs.harvard.edu/\#abs/2017A&A...597A..69S} {597, A69}

\bibitem[\protect\citeauthoryear{{Schaffer}, {Yang}  \& {Johansen}}{{Schaffer}
  et~al.}{2018}]{SYJ18}
{Schaffer} N.,  {Yang} C.-C.,   {Johansen} A.,  2018, \mn@doi [\aap]
  {10.1051/0004-6361/201832783}, \href
  {https://ui.adsabs.harvard.edu/abs/2018A&A...618A..75S} {618, A75}

\bibitem[\protect\citeauthoryear{{Surville} \& {Mayer}}{{Surville} \&
  {Mayer}}{2019}]{SM19}
{Surville} C.,  {Mayer} L.,  2019, \mn@doi [\apj] {10.3847/1538-4357/ab3e47},
  \href {https://ui.adsabs.harvard.edu/abs/2019ApJ...883..176S} {883, 176}

\bibitem[\protect\citeauthoryear{{Surville}, {Mayer}  \& {Lin}}{{Surville}
  et~al.}{2016}]{SML16}
{Surville} C.,  {Mayer} L.,   {Lin} D. N.~C.,  2016, \mn@doi [\apj]
  {10.3847/0004-637X/831/1/82}, \href
  {https://ui.adsabs.harvard.edu/abs/2016ApJ...831...82S} {831, 82}

\bibitem[\protect\citeauthoryear{{Tamura}}{{Tamura}}{2016}]{mT16}
{Tamura} M.,  2016, \mn@doi [Proceeding of the Japan Academy, Series B]
  {10.2183/pjab.92.45}, \href
  {https://ui.adsabs.harvard.edu/abs/2016PJAB...92...45T} {92, 45}

\bibitem[\protect\citeauthoryear{{Tanaka}, {Takeuchi}  \& {Ward}}{{Tanaka}
  et~al.}{2002}]{TTW02}
{Tanaka} H.,  {Takeuchi} T.,   {Ward} W.~R.,  2002, \mn@doi [\apj]
  {10.1086/324713}, \href
  {https://ui.adsabs.harvard.edu/abs/2002ApJ...565.1257T} {565, 1257}

\bibitem[\protect\citeauthoryear{{Yang} \& {Johansen}}{{Yang} \&
  {Johansen}}{2014}]{YJ14}
{Yang} C.-C.,  {Johansen} A.,  2014, \mn@doi [\apj]
  {10.1088/0004-637X/792/2/86}, \href
  {https://ui.adsabs.harvard.edu/abs/2014ApJ...792...86Y} {792, 86}

\bibitem[\protect\citeauthoryear{{Yang} \& {Johansen}}{{Yang} \&
  {Johansen}}{2016}]{YJ16}
{Yang} C.-C.,  {Johansen} A.,  2016, \mn@doi [\apjs]
  {10.3847/0067-0049/224/2/39}, \href
  {https://ui.adsabs.harvard.edu/#abs/2016ApJS..224...39Y} {224, 39}

\bibitem[\protect\citeauthoryear{{Yang} \& {Krumholz}}{{Yang} \&
  {Krumholz}}{2012}]{YK12}
{Yang} C.-C.,  {Krumholz} M.,  2012, \mn@doi [\apj]
  {10.1088/0004-637X/758/1/48}, \href
  {https://ui.adsabs.harvard.edu/#abs/2012ApJ...758...48Y} {758, 48}

\bibitem[\protect\citeauthoryear{{Yang} \& {Menou}}{{Yang} \&
  {Menou}}{2010}]{YM10}
{Yang} C.-C.,  {Menou} K.,  2010, \mn@doi [Monthly Notices of the Royal
  Astronomical Society] {10.1111/j.1365-2966.2009.16047.x}, \href
  {https://ui.adsabs.harvard.edu/abs/2010MNRAS.402.2436Y} {402, 2436}

\bibitem[\protect\citeauthoryear{{Yang}, {Johansen}  \& {Carrera}}{{Yang}
  et~al.}{2017}]{YJC17}
{Yang} C.~C.,  {Johansen} A.,   {Carrera} D.,  2017, \mn@doi [\aap]
  {10.1051/0004-6361/201630106}, \href
  {https://ui.adsabs.harvard.edu/\#abs/2017A&A...606A..80Y} {606, A80}

\bibitem[\protect\citeauthoryear{{Yang}, {Mac Low}  \& {Johansen}}{{Yang}
  et~al.}{2018}]{YMJ18}
{Yang} C.-C.,  {Mac Low} M.-M.,   {Johansen} A.,  2018, \mn@doi [\apj]
  {10.3847/1538-4357/aae7d4}, \href
  {https://ui.adsabs.harvard.edu/\#abs/2018ApJ...868...27Y} {868, 27}

\bibitem[\protect\citeauthoryear{{Youdin} \& {Goodman}}{{Youdin} \&
  {Goodman}}{2005}]{YG05}
{Youdin} A.~N.,  {Goodman} J.,  2005, \mn@doi [\apj] {10.1086/426895}, \href
  {https://ui.adsabs.harvard.edu/abs/2005ApJ...620..459Y} {620, 459}

\bibitem[\protect\citeauthoryear{{Youdin} \& {Lithwick}}{{Youdin} \&
  {Lithwick}}{2007}]{YL07}
{Youdin} A.~N.,  {Lithwick} Y.,  2007, \mn@doi [\icarus]
  {10.1016/j.icarus.2007.07.012}, \href
  {https://ui.adsabs.harvard.edu/abs/2007Icar..192..588Y} {192, 588}

\bibitem[\protect\citeauthoryear{{Zhang} \& {Zhu}}{{Zhang} \&
  {Zhu}}{2019}]{ZZ19}
{Zhang} S.,  {Zhu} Z.,  2019, \mnras, submitted (\href
  {https://arxiv.org/abs/1911.01530} {arXiv:1911.01530})

\bibitem[\protect\citeauthoryear{{Zhang} et~al.,}{{Zhang} et~al.}{2018}]{ZZ18}
{Zhang} S.,  et~al., 2018, \mn@doi [\apjl] {10.3847/2041-8213/aaf744}, \href
  {https://ui.adsabs.harvard.edu/abs/2018ApJ...869L..47Z} {869, L47}

\bibitem[\protect\citeauthoryear{{Zhu} \& {Stone}}{{Zhu} \&
  {Stone}}{2018}]{ZS18}
{Zhu} Z.,  {Stone} J.~M.,  2018, \mn@doi [\apj] {10.3847/1538-4357/aaafc9},
  \href {https://ui.adsabs.harvard.edu/abs/2018ApJ...857...34Z} {857, 34}

\bibitem[\protect\citeauthoryear{{Zhu}, {Nelson}, {Dong}, {Espaillat}  \&
  {Hartmann}}{{Zhu} et~al.}{2012a}]{ZN12}
{Zhu} Z.,  {Nelson} R.~P.,  {Dong} R.,  {Espaillat} C.,   {Hartmann} L.,
  2012a, \mn@doi [\apj] {10.1088/0004-637X/755/1/6}, \href
  {https://ui.adsabs.harvard.edu/abs/2012ApJ...755....6Z} {755, 6}

\bibitem[\protect\citeauthoryear{{Zhu}, {Stone}  \& {Rafikov}}{{Zhu}
  et~al.}{2012b}]{ZSR12}
{Zhu} Z.,  {Stone} J.~M.,   {Rafikov} R.~R.,  2012b, \mn@doi [\apj]
  {10.1088/2041-8205/758/2/L42}, \href
  {https://ui.adsabs.harvard.edu/\#abs/2012ApJ...758L..42Z} {758, L42}

\bibitem[\protect\citeauthoryear{{Zhu}, {Stone}  \& {Rafikov}}{{Zhu}
  et~al.}{2013}]{ZSR13}
{Zhu} Z.,  {Stone} J.~M.,   {Rafikov} R.~R.,  2013, \mn@doi [\apj]
  {10.1088/0004-637X/768/2/143}, \href
  {https://ui.adsabs.harvard.edu/abs/2013ApJ...768..143Z} {768, 143}

\bibitem[\protect\citeauthoryear{{Zhu}, {Stone}, {Rafikov}  \& {Bai}}{{Zhu}
  et~al.}{2014}]{ZS14}
{Zhu} Z.,  {Stone} J.~M.,  {Rafikov} R.~R.,   {Bai} X.-n.,  2014, \mn@doi
  [\apj] {10.1088/0004-637X/785/2/122}, \href
  {https://ui.adsabs.harvard.edu/abs/2014ApJ...785..122Z} {785, 122}

\bibitem[\protect\citeauthoryear{{Zhu}, {Stone}  \& {Bai}}{{Zhu}
  et~al.}{2015a}]{ZSB15}
{Zhu} Z.,  {Stone} J.~M.,   {Bai} X.-N.,  2015a, \mn@doi [\apj]
  {10.1088/0004-637X/801/2/81}, \href
  {https://ui.adsabs.harvard.edu/abs/2015ApJ...801...81Z} {801, 81}

\bibitem[\protect\citeauthoryear{{Zhu}, {Dong}, {Stone}  \& {Rafikov}}{{Zhu}
  et~al.}{2015b}]{ZD15}
{Zhu} Z.,  {Dong} R.,  {Stone} J.~M.,   {Rafikov} R.~R.,  2015b, \mn@doi [\apj]
  {10.1088/0004-637X/813/2/88}, \href
  {https://ui.adsabs.harvard.edu/abs/2015ApJ...813...88Z} {813, 88}

\bibitem[\protect\citeauthoryear{{de Val-Borro}, {Artymowicz}, {D'Angelo}  \&
  {Peplinski}}{{de Val-Borro} et~al.}{2007}]{VA07}
{de Val-Borro} M.,  {Artymowicz} P.,  {D'Angelo} G.,   {Peplinski} A.,  2007,
  \mn@doi [\aap] {10.1051/0004-6361:20077169}, \href
  {https://ui.adsabs.harvard.edu/abs/2007A&A...471.1043D} {471, 1043}

\bibitem[\protect\citeauthoryear{{van der Marel} et~al.,}{{van der Marel}
  et~al.}{2013}]{MD13}
{van der Marel} N.,  et~al., 2013, \mn@doi [Science] {10.1126/science.1236770},
  \href {https://ui.adsabs.harvard.edu/abs/2013Sci...340.1199V} {340, 1199}

\makeatother
\end{thebibliography}

%
%
\bsp	
\label{lastpage}
\end{document}